\documentclass{aa} 
\usepackage{graphicx}
\usepackage{lscape}
\usepackage{longtable}
\usepackage{arydshln}
\usepackage{multirow}
\usepackage{enumitem}
\usepackage{epstopdf}
\usepackage{natbib}

\bibpunct{(}{)}{;}{a}{}{,} 

\begin{document}
\title{\textit{Herschel}-PACS observations of far-IR lines in YSOs I: [OI] and $\rm H_{2}O$ at 63 $\rm \mu m$ \thanks{Tables B1, B2 and B3 are available in electronic form
at the CDS via anonymous ftp to cdsarc.u-strasbg.fr (130.79.128.5)
or via http://cdsweb.u-strasbg.fr/cgi-bin/qcat?J/A+A/} \thanks{{\it Herschel} is an ESA space observatory with
    science instruments provided by European-led Principal
    Investigator consortia and with important participation from
    NASA.}}
   \author{P. Riviere-Marichalar\inst{1}, B. Mer\'in\inst{1}, I. Kamp\inst{2}, C. Eiroa\inst{3}, B. Montesinos\inst{4} 
}

   \institute{European Space Astronomy Centre (ESA), P.O. Box 78, 28691 Villanueva de la Ca\~nada, Spain 
                  \email{priviere@sciops.esa.int}
   \and Kapteyn Astronomical Institute, University of Groningen, P.O. Box 800, 9700 AV Groningen, The Netherlands 
   \and Dep. de F\'isica Te\'orica, Fac. de Ciencias, UAM Campus Cantoblanco, 28049 Madrid, Spain 
   \and Depto. Astrof\'isica , Centro de Astrobiolog\'{\i}a (CAB, INTA--CSIC), P.O. Box 78, ESAC Campus, 28691 Villanueva de la Ca\~nada, Madrid, Spain 
   }
   \authorrunning{Riviere-Marichalar et al.}
   \date{}

 \abstract 
{Gas plays a major role in the dynamical evolution of young stellar objects (YSOs). Its interaction with the dust is the key to our understanding planet formation later on in the protoplanetary disc stage.  Studying the gas content is therefore a crucial step towards understanding YSO and planet formation. Such a study can be made through spectroscopic observations of emission lines in the far-infrared, where some of the most important gas coolants emit, such as the [OI] $\rm ^{3}P_{1} \rightarrow$$\rm ^{3}P_{2}$ transition at 63.18 $\rm \mu m$. }
{We provide a compilation of observations of far-IR lines in 362 YSOs covering all evolutionary stages, from Class 0 to Class III with debris discs. In the present paper we focus on [OI] and o-$\rm H_{2}O$ emission at 63 $\rm \mu m$.} 
{We retrieved all the available \textit{Herschel}-PACS spectroscopic observations at 63 $\rm \mu m$ that used the dominant observing mode, the chop-nod technique. We provide measurements of line fluxes for the [OI] $\rm ^{3}P_{1} \rightarrow$$\rm ^{3}P_{2}$ and o-$\rm H_{2}O$ $\rm 8_{08} \rightarrow 7_{17}$ transitions at 63 $\rm \mu m$ computed using different methods. Taking advantage of the PACS IFU, we checked for spatially extended emission and also studied multiple dynamical components in line emission.}
{The final compilation consists of line and continuum fluxes at 63 $\rm \mu m$ for a total of 362 young stellar objects (YSOs). We detect [OI] line emission at 63 $\rm \mu m$ in 194 sources out of 362, and line absorption in another five sources. o-$\rm H_{2}O$ was detected in 42 sources. We find evidence of extended [OI] emission in 77 sources, and detect 3$\sigma$ residual emission in 71 of them. The number of sources showing extended emission decays from Class 0 to Class II. We also searched for different components contributing to the line emission, and found evidence for multiple components in 30 sources. We explored correlations between line emission and continuum emission and found a clear correlation between WISE fluxes from 4.6 to 22 $\rm \mu m$ and [OI] line emission. We conclude that the observed emission is typically a combination of disc, envelope and jet emission.}  
{}
\keywords{Stars: Circumstellar matter, Stars: evolution, astrochemistry, protoplanetary disks}

   \maketitle

\section{Introduction} 
Young stellar objects (YSO) are complex sources consisting of many components, such as the central source (protostellar or stellar), an envelope made of gas and dust, a circumstellar disc, stellar and disc winds, and large-scale collimated jets. Each of the components can contribute to different observables, such as photometry and line fluxes. A detailed study is therefore needed to elucidate the contribution of each component. 

In the initial stages of stellar formation, Class 0 and I protostars \citep{Lada1984,Lada1987,Andre1993}  are surrounded by an envelope. Discs are clearly detected around Class I sources. Class I sources later evolve to Class II sources, in which the central star is already formed and the envelope dispersed. The formation of a dust opacity hole in the inner disc leads to the formation of the so-called transitional discs \citep{Strom1989}. Many mechanisms have been used to explain the formation of the inner opacity holes, including planet formation. At 10  Myr, most primordial discs have been dispersed \citep{Strom1989}, but destructive collisions between planetesimals can repopulate the circumstellar environment with dust, resulting in the so-called debris discs. 

Young stellar objects can also be classified according to their masses. The so-called T Tauri stars are variable stars showing bright emission lines with stellar masses $\rm M_{*} < 2.0 M_{\odot}$, while HAeBe stars are the high-mass counterparts of T Tauri stars ($\rm 2.0 < M/M_{\odot} < 8.0$). 

Although gas is thought to dominate the mass budget during the primordial stages (Class 0 to II), little is known about its mass and spatial distribution, mostly because it is difficult to detect $\rm H_{2}$, which lacks a permanent dipole moment. However, to learn about the formation of planets, we need to understand the chemical evolution of gas and dust. 

The \textit{Herschel Space Observatory} \citep{Pilbratt2010} produced thousands of observations of YSOs during its four-year mission. The most widely used instrument was the Photodetector Array Camera and Spectrometer \citep[PACS,][]{Poglitsch2010}, which can spectroscopically observe the far-IR 50-250 $\rm \mu m$ range. Furthermore, it also performed photometric observations at 70, 100 and 160 $\rm \mu m$ with great sensitivity. One of the most interesting characteristics of the PACS spectrometer is its Integral Field Unit (IFU), divided into 25 spaxels distributed in a regular grid covering 47\arcsec $\rm \times$47\arcsec. The IFU allows us to study the spatial distribution of the continuum and line emission. 

Some studies have surveyed [OI], CO, OH, and $\rm H_{2}O$ emission in objects belonging to different stellar associations and moving groups using \textit{Herschel} \citep{Donaldson2012,Howard2013,Green2013,Mathews2013,Lindberg2014,Riviere2013,Riviere2014,Riviere2015}. Other studies have focused on the analysis of individual sources \citep{Meeus2010, vanKempen2010, Kempen2010, Sturm2010, Thi2010, Tilling2012, Lebreton2012, Riviere2012b, Thi2013}. However,  the spatial extension of the emission was discussed in only a few cases \citep[][]{Karska2013,Karska2014B,Nisini2013,Nisini2015}. The most extensively studied wavelength range is 63.0-63.4 $\rm \mu m$, which includes two transitions,  [OI] $\rm ^{3}P_{1} \rightarrow $$\rm^{3}P_{2}$ at 63.185 $\rm \mu m$  and  o-$\rm H_{2}O$ $\rm 8_{08} \rightarrow 7_{17}$ at 63.325 $\rm \mu m$. [OI] emission has been detected in YSOs at all evolutionary stages, from Class 0 and I \citep{Green2013} to Class II and transitional \citep{Howard2013} and  debris discs \citep{Riviere2012}. o-$\rm H_{2}O$ emission was observed around Class 0, I, II and transition discs, but not around debris discs.

Understanding the spatial distribution of far-IR lines emission is crucial, since it has been shown that envelopes, protoplanetary discs, and outflows can contribute to [OI] emission \citep{vanKempen2010,Podio2012, Karska2013}. [OI] extended emission along the jet direction has been commonly observed, while molecular extended emission is observed in only a few cases \citep{vanKempen2010,Herczeg2012}. \cite{Podio2012} and \cite{Karska2013} explained the extended emission as being produced by J- and C-shocks along the jet, and noted a decay in far-IR lines intensity from Class 0/I to Class II. \cite{Howard2013} studied a sample of Class II sources in Taurus, including sources with and without a jet or an outflow. The authors found a tight correlation between continuum emission at 63 $\rm \mu m$ and [OI] emission, suggesting a disc origin for the line. However, sources with jets show a brighter [OI] emission for the same level of continuum, indicating a contribution from the jet. The authors did not find a correlation between disc mass (derived from sub-millimeter continuum emission) and [OI] line intensity, indicating that either the line is optically thick or it is a poor tracer of gas mass. \cite{Green2013} studied a sample of 30 embedded sources (Class 0 and I) from the DIGIT program \citep[see e.g.][]{Kempen2010,Sturm2010} and found a tight correlation between line intensity and $\rm L_{bol}$.

In this paper, we present a compilation  of 432 PACS spectroscopic observations of 362 YSOs and main-sequence stars with debris discs. We focus on the small wavelength range between 63.0 and 63.4 $\rm \mu m$, which includes the [OI] transition at 63.185 $\rm \mu m$ and the o-$\rm H_{2}O$ transition at 63.325 $\rm \mu m$. Our wavelength range selection is motivated by the fact that the [OI] transition at 63.185 $\rm \mu m$ typically is the strongest line coolant in protoplanetary discs \citep{Gorti2008}. We leave the study of other transitions observed in \textit{PACS} range mode for a future paper. 

\section{Sample and observations}\label{sec:Sample}
\begin{figure}[!t]
\begin{center}
 \includegraphics[]{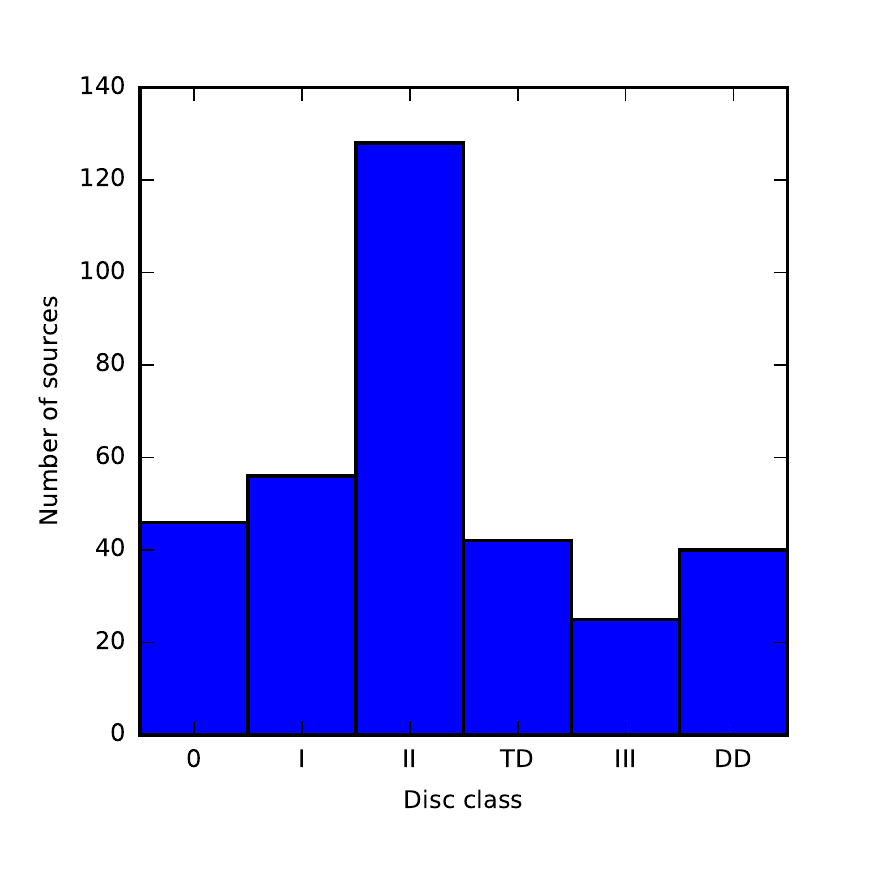}
\caption{Distribution of evolutionary stages in the observed sample.}
\label{Fig:discc_hist}
\end{center}
\end{figure}

The data were collected from fourteen different programs (see Table \ref{Tab:programs}).  The sample consists of 362 YSOs and main-sequence stars with debris discs observed with PACS spectroscopy at 63 $\rm \mu m$. The total number of observations was 432: 51 sources were observed twice, and another nine sources were observed three times. We only included pointed observations from programs that used the chop-nod technique to remove the telescope and background contribution. The sample includes objects in all the different evolutionary stages for circumstellar material, from Class 0 to Class III stars and to those that are later surrounded by debris discs (see Fig. \ref{Fig:discc_hist}). Highly embedded sources from the DIGIT program \citep{Sturm2010} were included, together with highly embedded and high-mass protostellar envelopes from the WISH program \citep{Dishoeck2011}. Transition discs are a particular case of Class II discs that have opened a gap in the inner part of the disc. 

Throughout the paper we treat transition discs separately because of their importance in testing planet formation theories. When we refer to Class II discs, we therefore refer to full discs without an inner gap, in contrast to transition discs which show inner opacity holes. Because of the limited size of the samples and because of the similarities we found, we treat Class 0 and I sources in some sections as a single group of protostellar objects (see Sect. \ref{Sec:results}). Most of the stars in the sample belong to different star-forming regions and stellar associations, with ages in the range 1--40 Myr, including Tau, Cha, Cha II, Lupus, Lupus III,  $\rm \eta Cha$, Upper Scorpius, TWA, BPMG, Tuc Hor, CrA, Serpens, Per, and Oph. The source names, positions, spectral type (for Class II and III sources), evolutionary status, and associations for YSOs in the sample are given in Table \ref{YSO_sample}.

\begin{table}[!t]
\caption{Overview of the programs used in this study.}             
\label{Tab:programs}      
\centering          
\begin{tabular}{lll}     
\hline \hline 
Program ID & observations & Sensitivity\\
-- & -- & $\rm (10^{-18}W/m^{2})$ \\
\hline
\textit{GT1\_vgeers\_1} & 2 & 2.5 \\
\textit{KPOT\_bdent\_1} & 185 & 3.1 \\
\textit{KPOT\_nevans\_1} & 65 & 7.3 \\
\textit{KPGT\_golofs01\_1}  & 6 & 2.9 \\
\textit{KPGT\_evandish\_1}  & 28 & 8.6 \\
\textit{OT1\_ascholz\_1} & 1 &  1.2 \\
\textit{OT1\_cespaill\_2} & 38 & 20.0 \\
\textit{OT1\_ckiss\_1} & 9 & 3.5 \\
\textit{OT1\_gmeeus\_1} & 2 & 28.0 \\
\textit{OT1\_ipascucc\_1} & 30 & 1.6 \\
\textit{OT1\_maudar01\_1} & 11 & 23.0 \\
\textit{OT1\_vgeers\_2} & 4 & 2.5 \\
\textit{OT2\_amoor\_3} & 2 & 3.8 \\
\textit{OT2\_evandish\_4} & 49 & 7.7 \\
\hline                  
\end{tabular}
\end{table}

Observations were performed in two different modes depending on the program: line spectroscopy, covering 63.0 to 63.4  $\rm  \mu m$ (301 observations), and range spectroscopy (131 observations), covering 55 to 72 $\rm \mu m$. Line spectroscopy observes a narrow spectral region centred on a certain spectroscopic transition ( [OI] $\rm ^{3}P_{1} \rightarrow$$\rm ^{3}P_{2}$ in our case), and guarantees the detection of the full line profile for an unresolved line, with enough continuum coverage at either side of the line to allow for continuum measurements. Range spectroscopy covers a region around the lines of interest defined by the observer. The 63.0-63.4 $\rm \mu m$ range contains the [OI] $\rm ^{3}P_{1} \rightarrow$$\rm^{3}P_{2}$ transition at 63.185 $\rm \mu m$ and the o-$\rm H_{2}O$ $\rm 8_{18} \rightarrow 7_{07}$ transition at 63.325 $\rm \mu m$. We focus in this paper on the study of the two lines in the 63 to 63.4 $\rm \mu m$ wavelength range. Exposure times ranged from 851 s to 16420 s, with 75\% of the observations having $\rm t_{exp}<4000~s$. All the observations were performed in chop-nod mode, with a small chopper throw (1$\rm \arcmin$.5) for 299 sources ($\rm \sim$71\%), medium chopper throw (3$\rm \arcmin$) for 13 ($\rm \sim$3\%) and large chopper throw (6$\rm \arcmin$) for 107 ($\rm \sim$26\%).   Most observations used only one nod cycle ($\rm \sim$67\%), $\rm \sim$26 \% used two nod cycles, four observations ($\rm \sim$1\%) used three nod cycles,  23 ($\rm \sim$5\%) used four nod cycles and six used five nod cycles (1\%).

\begin{figure*}[!t]
\begin{center}
\includegraphics[]{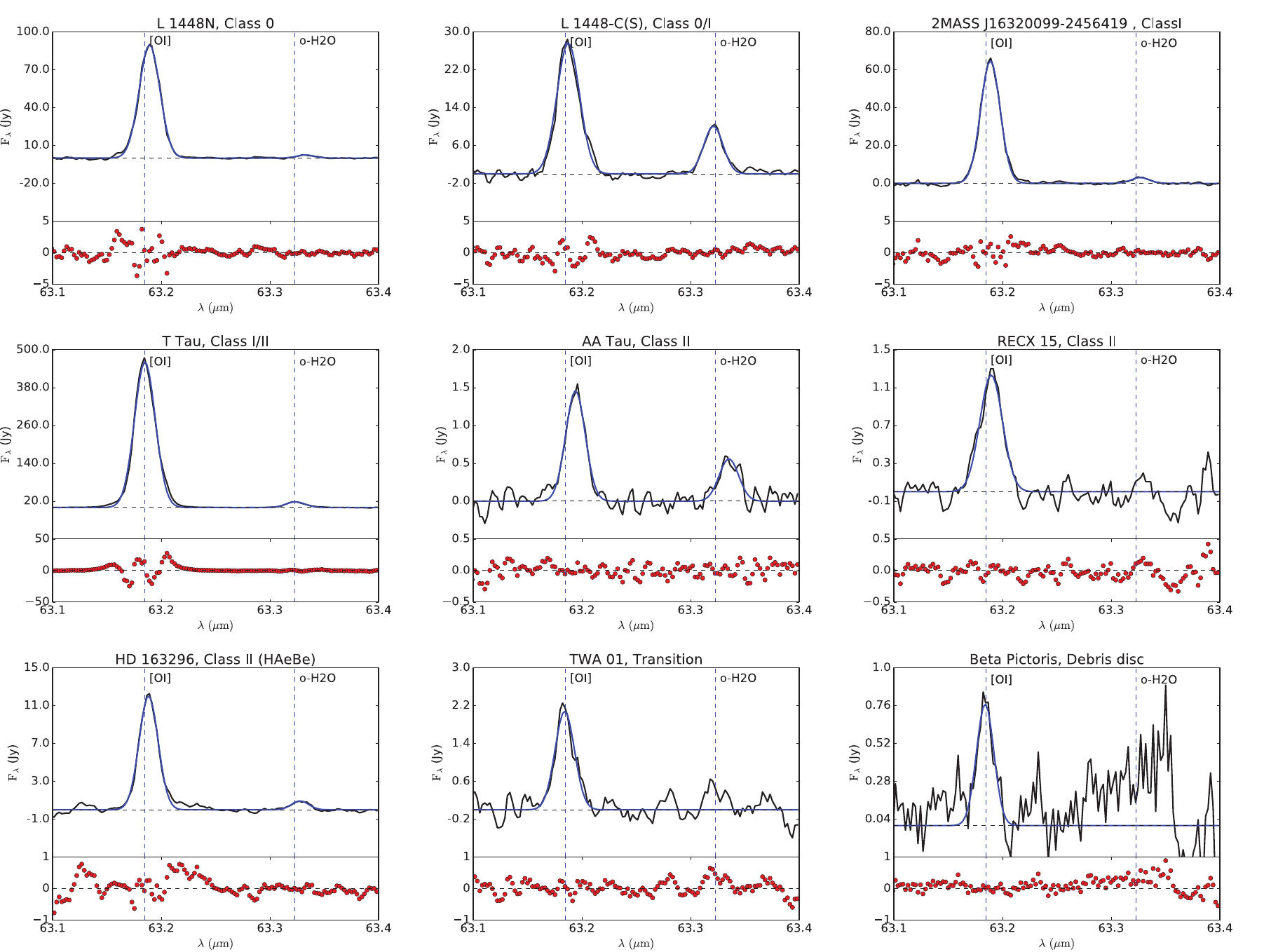}\
\caption{Example of continuum-subtracted spectra of stars at different evolutionary stages. The black line shows the observed spectra, while the blue line shows a Gaussian fit. We show the rest-frame wavelength of the [OI] and o-$\rm H_{2}O$ emission lines and vertical dashed blue lines with labels identifying each line. Plots with the model residuals are shown at the bottom of each spectrum. }
\label{Fig:ExampleSp}
\end{center}
\end{figure*}

\section{Data reduction}\label{sec:obsDat}
The data were reduced using HIPE 12.0. The reduction was performed as follows. First, the observations were corrected for satellite movements. Then, saturated  frames and frames with glitches were flagged and masked, and the chop-off position was subtracted from the chop-on position to remove the sky and telescope contribution. Then, the cubes were divided by the spectral response function, and after that, flat-fielding, using a straight line, was applied to improve the signal-to-noise ratio (S/N) of the continuum. We used a straight line for flat-fielding. Then, the spectra were rebinned, with oversample = 2 and upsample =4. Observations from the \textit{OT1\_cespaill\_2} and \textit{OT1\_maudar01\_1} programs, and some observations from the program \textit{KPOT\_nevans\_1}, required oversample=2 and upsample=2 (native resolution of the instrument). Finally, the mean value from the two nod positions was computed. 

Given the noise increment in the edges of the spectra, we only considered the spectral range from 63.0 to 63.4 $\rm \mu m$ for line spectra and the range 62.5-63.9 $\rm \mu m$ for range spectra. We subtracted the continuum contribution by fitting a first-order polynomial after masking a $\rm \pm 3\sigma $ (where $\sigma$ is the width of an unresolved line at the wavelength of interest) region around each transition present ([OI] and o-$\rm H_{2}O$ at 63.185 and 63.325 $\rm \mu m$, respectively). Line fluxes were computed by integrating a Gaussian fit to continuum-subtracted spectra of the spaxel with the highest flux. When the continuum was not detected to the 3$\sigma$ level, we extracted the spectrum from the central spaxel (x=2, y=2). The spectrum from this spaxel was then aperture corrected. For some sources, a mis-pointing of \textit{Herschel} causes the source to lie in between many spaxels, and therefore the computed flux is a lower limit to the actual flux. Adding the 3$\rm \times$3 spaxels around the position of the source gives a more accurate flux for these sources. The  uncertainties on the line fluxes were computed as the integral of a Gaussian with a width equal to the fitted value and a peak equal to the noise of the continuum. Three-sigma upper limits were computed in a similar way as three times the integral of a Gaussian with a width equal to the instrumental value and a peak equal to the noise of the continuum.  Given the spread in observing times between the different programs, the sensitivity limits are in the range 1.2$\rm \times 10^{-18}$ to 2.8$\rm \times 10^{-17}~W/m^{2}$ (see Table \ref{Tab:programs}). We also obtained continuum fluxes at 63 $\rm \mu m$ by computing the mean value of the baseline after excluding the 3$\sigma$ regions around the position of detectable lines, with errors being the standard deviation inside the same region.

\section{Results}\label{Sec:results}

\begin{figure}[!t]
\begin{center}
\includegraphics[]{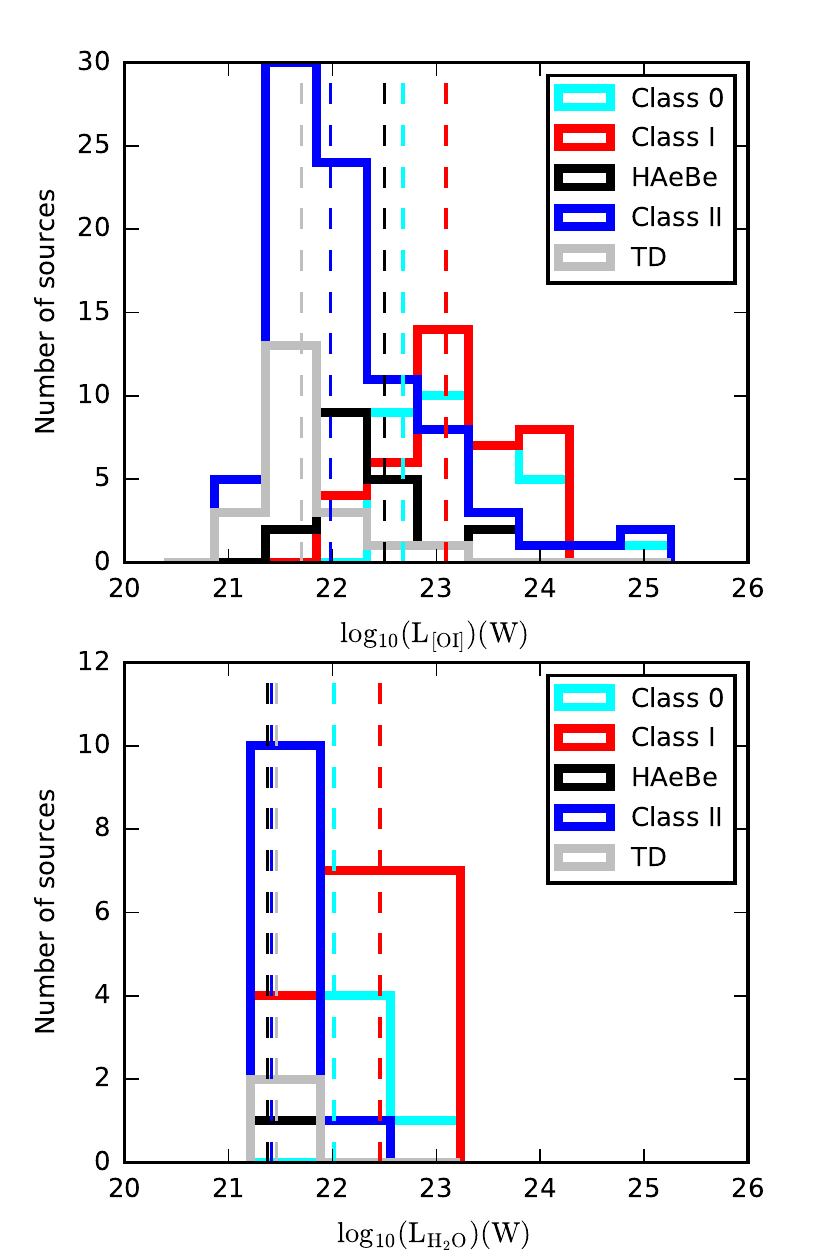}\\
  \caption{Top: histogram for [OI] line luminosities in the observed sample.  Bottom: histogram for $\rm H_{2}O$ line luminosities in the observed sample. The vertical dashed lines show the median luminosity for each class.}
   \label{Fig:hist_fOI}
\end{center}
\end{figure}

Examples of continuum-subtracted spectra for different evolutionary stages are shown in Fig. \ref{Fig:ExampleSp}. We also show the Gaussian fits to the observed profiles used to compute line fluxes, together with residual plots at the bottom of each spectrum. The shift in the observed line centres compared to the theoretical ones in some sources might be due to mis-pointing of the telescope, since it is a known PACS effect that telescope mis-pointing results in a shift in wavelengths, which hinders concluding wether the line shift is real. The strength of the [OI] line compared to the o-$\rm H_{2}O$ is evident from the plots, with typical line ratios $\rm F_{[OI]}/F_{H_{2}O}$ in the range 2.4 to 29. The only exception is BP Tau, where the o-$\rm H_{2}O$ and [OI] line fluxes at 63 $\rm \mu m$ are similar, within the errors.

\subsection{Line emission}
We detected the [OI]  emission line  at 63.185 $\rm \mu m$ (3$\rm \sigma$) in 194 sources out of 362 observed ($\rm 0.54 \pm 0.04$ detection fraction). Line fluxes from the central spaxel are given in Table \ref{YSO_fluxes}. The detection fractions strongly depend on the evolutionary stage. Class 0 and I sources show very similar detection fractions ($\rm 0.87^{+0.04}_{-0.07}$ and $\rm 0.93^{+0.02}_{-0.06}$, respectively), but show strong differences with Class II stars ($\rm 0.53^{+0.06}_{-0.05}$, including both T Tauri and HAeBe stars). The detection fraction, also seems to depend on the stellar mass, since there is a strong difference between T Tauri and HAeBe stars ($\rm 0.42^{+0.05}_{-0.05}$ and $\rm 0.96^{+0.01}_{-0.08}$, respectively). Transition discs and full Class II disc sources show detection fractions that are almost compatible ($\rm 0.56^{+0.07}_{-0.08}$ for transition discs). Finally, [OI] was detected towards four debris disc sources (HD 172555, $\beta$ Pictoris, RXJ18523-3700 and HD 141569), leading to the smallest detection fraction ($\rm 0.10^{+0.07}_{-0.03}$). The detection fraction is likely to be even smaller, since HD 141569 is considered either a HAeBe star \citep{Mendigutia2011} or a debris disc \citep{Marsh2002}.

o-$\rm H_{2}O$ at 63.325 $\rm \mu m$ was  detected in 43 out of 362 sources ($\rm 0.12 \pm 0.02$ detection fraction). o-$\rm H_{2}O$ line emission was only detected in sources where we also detected [OI] emission. Eight of these are Class 0 sources, two are intermediate Class 0/I sources, 20 are Class I sources, one (T Tau) is an intermediate Class I/II source, ten are Class II sources (eight of them are T Tauri stars, and two are HAeBe stars), and two are transitional discs.

[OI] line fluxes extracted from the central spaxel range from 4$\rm \times 10^{-18}$ to 4$\rm \times 10^{-14}$ $\rm W/m^{2}$ and o-$\rm H_{2}O$ line fluxes range from 6$\rm \times 10^{-18}$ to 7$\rm \times 10^{-16}$ $\rm W/m^{2}$. We show in Fig. \ref{Fig:hist_fOI} the distribution of [OI] (top) and $\rm H_{2}O$ (bottom) luminosities for the different types of sources. Class 0 and I sources show a similar distribution of [OI] luminosities. However, they show quite different $\rm H_{2}O$ distributions. Class II  and transitional discs show more differences in their [OI] luminosity distributions than they do in their $\rm H_{2}O$ luminosity distributions. The low number of $\rm H_{2}O$ detections precludes any further comparison of the distributions. The vertical dashed lines in  Fig. \ref{Fig:hist_fOI} show the median [OI] line luminosity for each class. We observe a clear evolutionary trend, with [OI] line luminosities that decrease from Class 0 to Class II and transitional, in agreement with the finding by \cite{Podio2012} and \cite{Karska2013}, where a decrease in molecular luminosity and total line luminosity was observed.
 
To test whether the differences of [OI] luminosity distributions between the different classes are real, we performed two-distribution Kolmogorov-Smirnov tests for the different pairs of datasets. Class 0 and I sources do not differ in their distribution of [OI] fluxes (P=0.93). As we find no difference between both distributions, we combined the two distributions for an additional comparison with the other classes.  The comparison with the distributions for Class II bear strong differences: P$\rm \ll 10^{-3}$ for Class II sources compared to Class 0 and I sources, where Class II includes full disc T Tauris, transition discs T Tauris and HAeBe stars. The difference is dominated by low-mass stars, since the probability is P$\rm  \ll 10^{-3}$ when HAeBe stars are excluded from the comparison. The comparison between T Tauri stars and HAeBe stars also shows strong differences (P$\rm \sim 10^{-3}$). We furthermore compared the distribution of luminosities for T Tauri stars surrounded by full discs and those surrounded by transitional discs, and found a probability P$\rm \sim 10^{-3}$ that both populations are build from the same distribution.We did not perform the comparison for debris discs sources due to the small number of detections. However, we highlight the large dispersion of [OI] fluxes observed in debris discs, with only four detections covering more than two orders of magnitude. Figure \ref{Fig:hist_fOI} also shows an evolution of the typical [OI] luminosity with class, with the peak of the distribution having its maximum at fainter luminosities when moving from Class 0 to Class II and transitional discs. 

\begin{figure*}[!t]
\begin{center}
\includegraphics[]{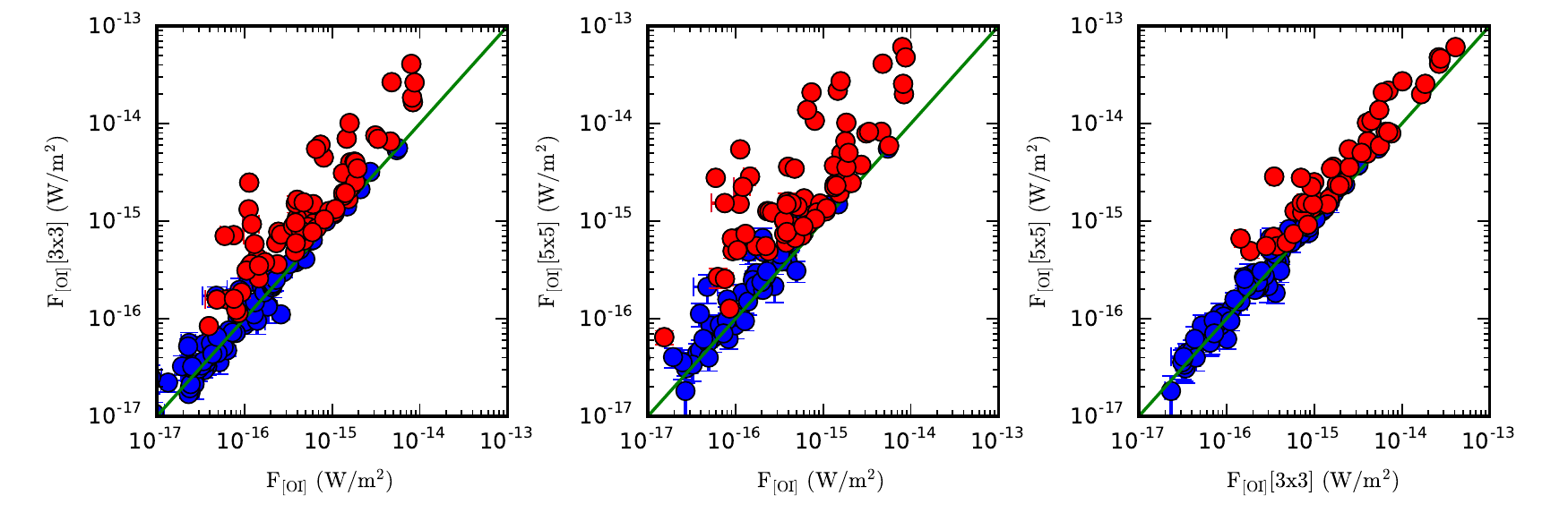}
   \caption{Extended emission tests for sources in the sample. Red dots identify sources showing extended emission in each of the tests. The solid diagonal line depicts a one-to-one ratio to help identify extended emission.}
   \label{Fig:ExtEm}
\end{center}
\end{figure*}

\begin{figure*}[!t]
\begin{center}
\includegraphics[]{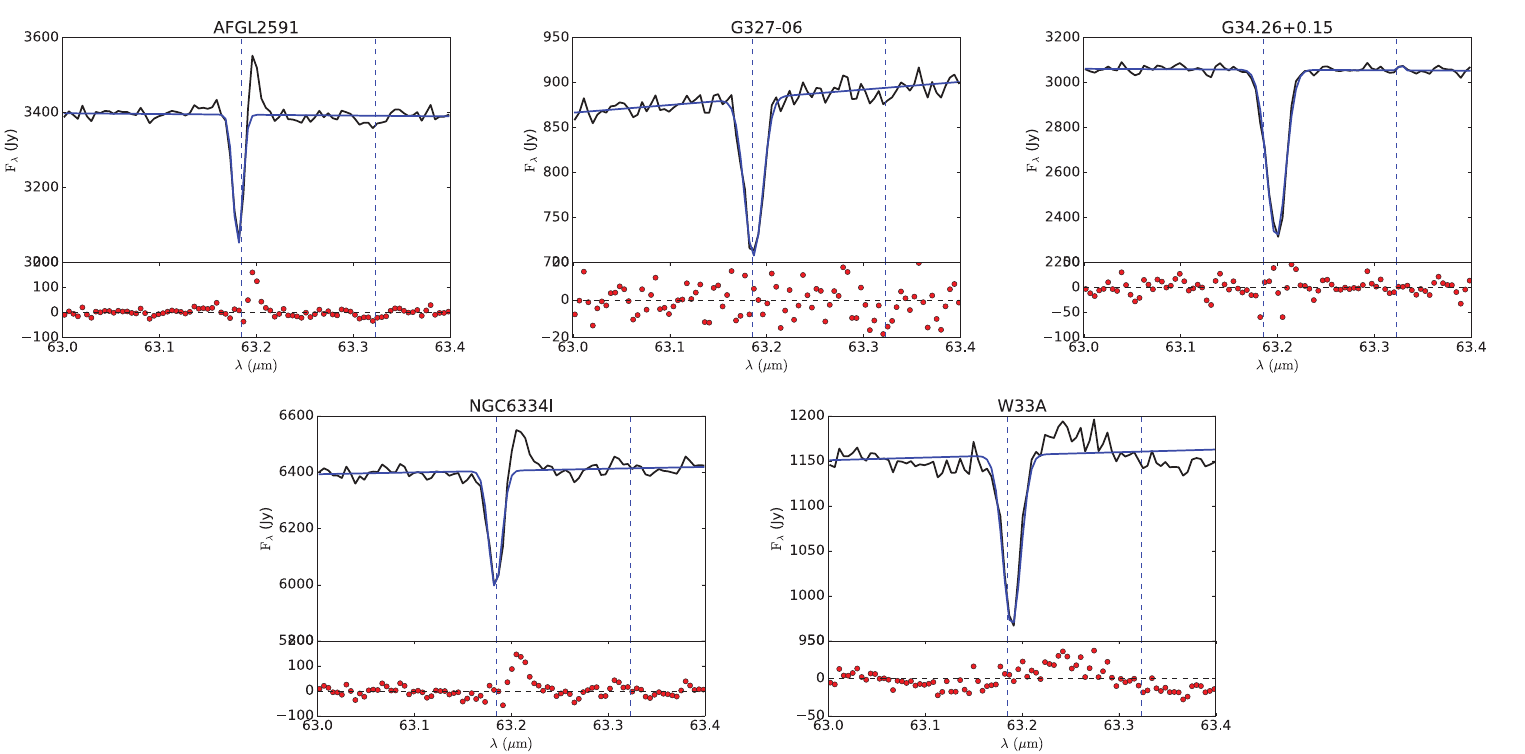}
   \caption{Spectral profiles for high-mass YSOs showing [OI] line absorption.}
   \label{Fig:Absorption}
\end{center}
\end{figure*}

\begin{figure*}[!t]
\begin{center}
\includegraphics[]{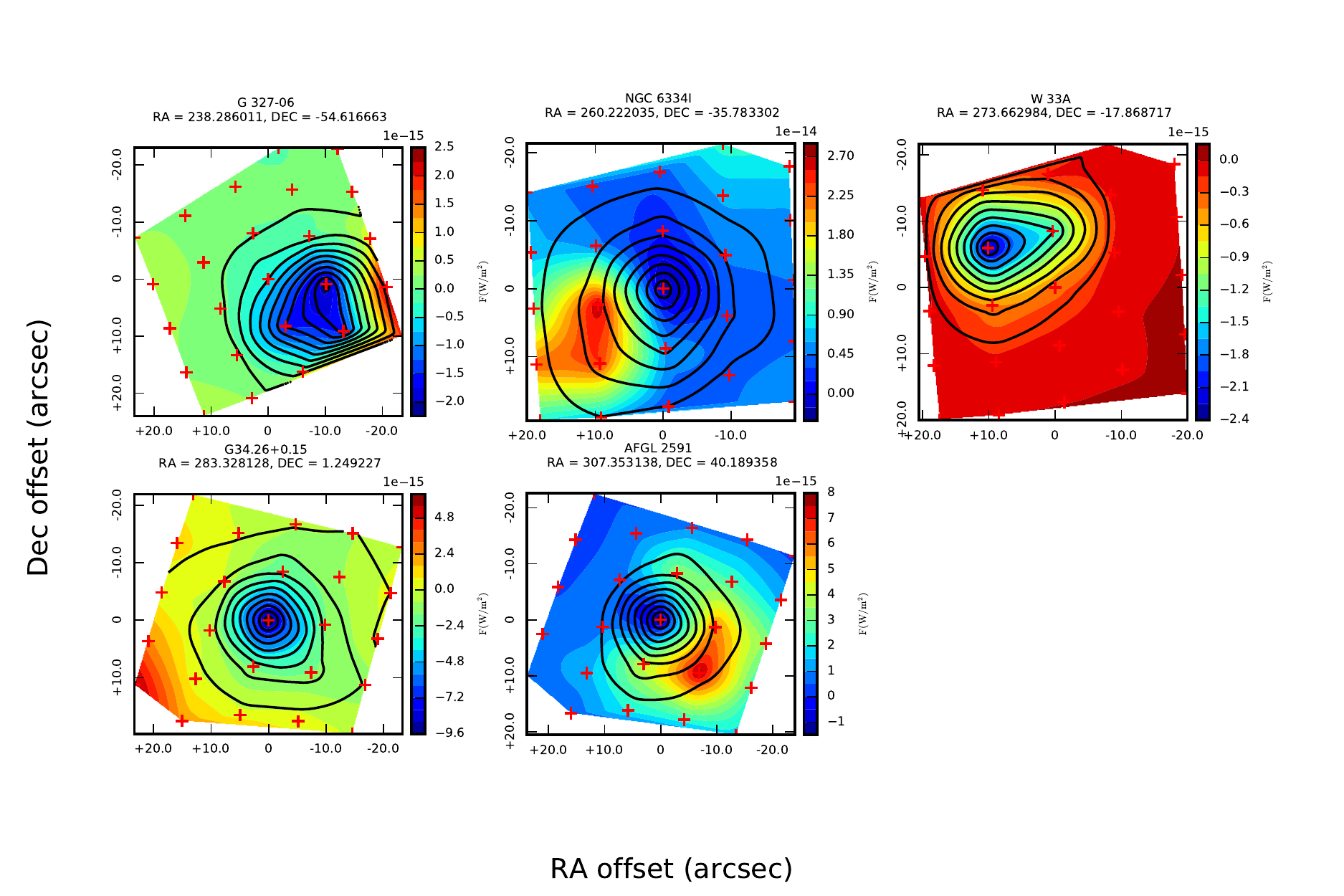}
 \caption{Line emission (coloured contours) and continuum maps at 63 $\rm \mu m$ (solid lines) for sources showing [OI] absorption at 63.185 $\rm \mu m$. The crosses mark the position of individual spaxels. }
   \label{Fig:AbsorptionMaps}
\end{center}
\end{figure*}

It is important to note that many sources suffer from mis-pointing, affecting both the fluxes and the future assessment of extended emission (Sect. \ref{Sec:ExtEmTest}). This mis-pointing can be in the form of small shifts, which makes the flux from the central spaxel a poor estimate of the real value. In other cases, the shift can cause the source to be located in a spaxel other than the central one. In these cases, the sources can again be properly centred inside this spaxel or it can be shifted. These sources are marked in Table  \ref{YSO_fluxes} with an asterisk. 

\subsection{Line absorption}
Five sources in the sample showed line absorption, at least in the central spaxel. These sources are AFGL 2591, G327-0.6, G34.26+0.15, NGC 6334-I, and W33A. Their spectra are shown in Fig. \ref{Fig:Absorption}. They are part of the WISH sample, and are classified as high-mass YSOs \citep{Dishoeck2011}. Since the observations were performed using the chop-nod technique, we inspected the on and off positions separately. We always found absorption in the on position, and only in one case, W33A, we observe faint emission in the off position. Therefore, we conclude that the observed absorption features are real. 

The sources showing absorption were discussed in detail by \cite{Karska2014}. For two sources, namely AFGL 2591 and NGC 6334-I, the profile features the shape of a P-Cygni profile \citep{Karska2014}. Line maps for sources showing absorption in the [OI] transition at 63.185 $\rm \mu m$ are shown in Fig. \ref{Fig:AbsorptionMaps}, where we can see that absorption peaks at the positions were the  continuum reaches the maximum. Absorption towards NGC 6334-V was previously reported by \cite{Kraemer1998}, who attributed the absorption to cooler or less dense gas in the foreground core cloud.

Even if all the sources showing line absorption were high-mass YSOs, we cannot conclude that high-mass envelopes lead to absorption, since another three high-mass YSOs in the sample, namely DR 21 (OH), NGC 7538 IRS1, and W3-IRS5 show prominent emission. None of the eight sources showed o-$\rm H_{2}O$ emission or absorption at 63.323 $\rm \mu m$. 

\subsection{Continuum and line variability}
\begin{figure*}[!t]
\begin{center}
\includegraphics[]{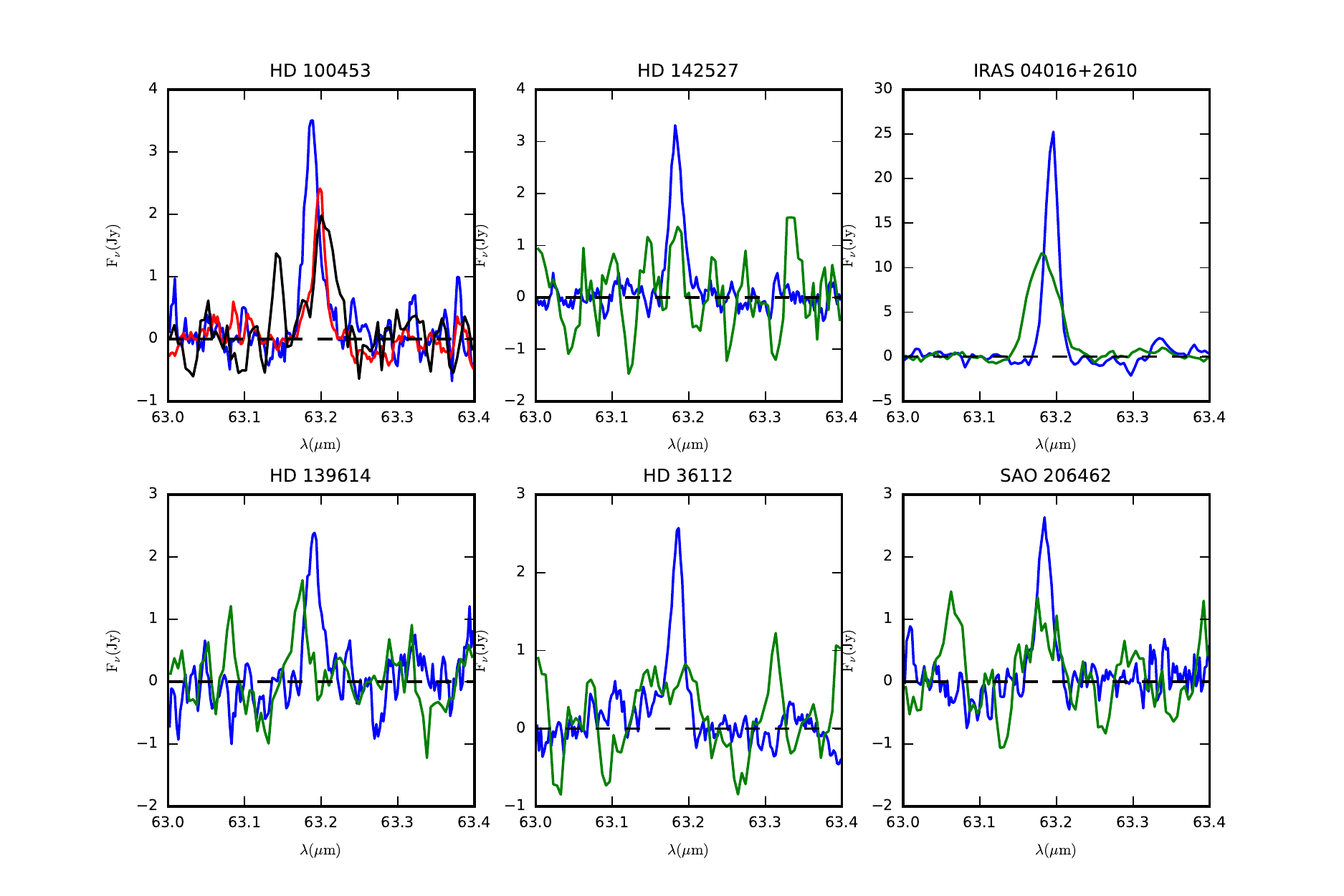}   
\caption{Spectra obtained at different epochs showing hints of line variability.}
   \label{Fig:variableSp}
\end{center}
\end{figure*}

A total of 60 sources were observed multiple times (see Sect. \ref{sec:Sample}, the discussed fluxes are shown in Table \ref{YSO_fluxes}). We can use this subsample to gain insight on line and continuum variability in the far-IR. In terms of line emission, we can distinguish three groups: 
\begin{itemize}
	\item Sources that show fluxes that agree, within the uncertainties. This group contains 38 sources.
	\item Sources that show fluxes that do not match becuase of a mis-pointing in one (or all) the observations. This groups contains 16 sources. Mis-pointed sources mostly come from the GASPS program, where a problem with the pointing of Taurus sources has been highlighted by \cite{Howard2013}. These sources are properly flagged in Table \ref{YSO_fluxes}.
	\item Sources with fluxes that do not match, and where the disagreement is not due to mis-pointing, but most likely to real variability or unknown instrumental effects. This group includes six sources: HD 100453, HD 139614, HD 142527, HD 36112, IRAS 04016+2610, and SAO 206462. The spectra obtained at different epochs for these sources are shown in Fig. \ref{Fig:variableSp}. Five of these (HD 100453, HD 139614, HD 142527, HD 36112 and SAO 206462) are HAeBe stars,  and one (IRAS 04016+2610) is a Class 0/I embedded source.  For IRAS 04016+2610 the difference in the fluxes could be due to the different observational techniques used, since one of the observations is a line spectrum, while the other is a range spectrum. Furthermore, all sources are variable at most at the 3$\sigma$ level. For HD 100453, three observations are available, and none of them shows compatible line fluxes at the 1$\sigma$ level, but they are compatible at the 3$\sigma$ level. Furthermore, the continuum sources are not compatible at either the 1 or 3$\sigma$ levels.
\end{itemize}

Follow-up of [OI] line emission at 63 $\rm \mu m$ is needed to understand line variability and rule out instrumental effects.

\subsection{Extended emission}\label{Sec:ExtEmTest}

The Integral Field Unit (IFU) used for PACS spectroscopy allowed us to study whether the emission is extended or not. We first compared the flux derived from the central spaxel (or the spaxel with the highest signal) to those derived from co-adding the nine central spaxels (test 1) and the 25 spaxels (test 2). Finally, we also compared the flux from the central nine spaxels with that of the 25 spaxels (test 3). If the emission is extended, we expect the co-added fluxes to be substantially higher than the flux from the central spaxel. This stronger emission from co-added spaxels can be due to extended emission or to the presence of multiple sources. On the one hand, if the flux from the 25 spaxels were higher than the flux from the central spaxel, but the flux from the central were  coincident with the flux from the central spaxel, then the most likely explanation would be the presence of another source (or sources) in the outer spaxels. On the other hand, if we were to detect higher fluxes only when the central nine spaxels are considered, then the difference would most likely be due to extended emission. We are aware that by co-adding the different spaxels, the signal detected in one of them can be diluted when the other spaxels are noise-dominated. 

\begin{figure}[!t]
\begin{center}
\includegraphics[]{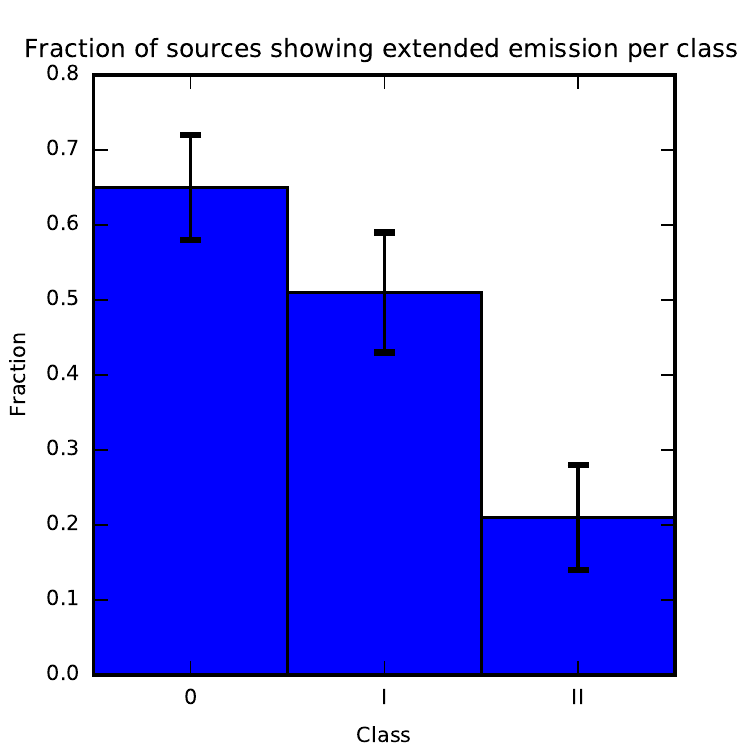}
   \caption{Bar plot showing the global fraction of sources with extended [OI] emission per class.}
   \label{Fig:histExt}
\end{center}
\end{figure}

We show in Table \ref{Tab:YSO_fluxes_ext} the computed  fluxes for sources detected with at least one of the two methods. In Fig. \ref{Fig:ExtEm}, we compare the different fluxes computed to test for extended emission. When the difference in flux is larger than three times the quadratic sum of the uncertainties, we consider that the emission is extended. Sources with extended emission are shown as red dots in Fig. \ref{Fig:ExtEm}. When test 1 was used, 69 sources showed extended emission, 59 of which belonged to Class 0 and I, 9 belonging to Class II and transitional (including both T Tauri and HAeBe stars) and one is a highly embedded source with unknown class  (RCrA-IRS7A). When test 2 was used, 69 sources showed extended emission, 56 of which belonged to Class 0 and I, 12 to Class II and transitional, and one is a highly embedded protostar  (RCrA-IRS7A). Finally, 51 sources showed extended emission when Test 3 was used, 41 of which belonged to Class 0 and I, eight belonged to Class II and transitional, and two are highly embedded protostars (RCrA-IRS7A and AFGL 2591, which shows absorption in the central spaxel, together with extended emission in the surrounding spaxels). The final fraction of extended sources per class is shown in Fig. \ref{Fig:histExt}.

For all the sources showing higher fluxes from co-added spaxels, we used the method by \cite{Podio2012} to detect extended emission. This method compares the ratio of line to continuum emission in the different spaxels with that in the central one, aiming to detect residual emission over the expected value. The method implicitly assumes  that the continuum is emitted by a point source. While this is true for most Class II sources and transitional discs, it might not be true for Class 0 and I sources in the sample, as shown by \cite{Lindberg2014}.

We detected 3$\sigma$ residual emission in the maps of 71 sources (line emission maps for these sources are shown in Fig. \ref{Fig:lineMaps} and residual maps are shown in Fig. \ref{Fig:ResidualMaps}). FS Tau and FS Tau B are both included in the same OBSID, making the algorithm identify false residual emission. The residual found westward of the source is indeed emitted by FS TauB. The fields for RCrA-IRS7B are too crowded, and therefore we decided to exclude them from the analysis. \cite{Lindberg2014} performed a detailed study of this region, and we refer to this paper for more detailed results.

To compare between continuum and line emission, we distinguished three groups of maps. In the first group, we included sources that showed the peaks of line and continuum emission at the same position. In total, 39 sources were included in this first group. Prominent examples of this type are the maps of sources such as IRAS 03235+3004 or IRAS 04264+2433. In the second group, we included 18 maps where continuum and line emission peak at different positions. Examples of this sources are NGC1333 IRAS 4A and VLA 1623-243. Finally, the third group, with 14 observations, included very complex maps, mostly due to the presence of multiple sources. For this third group, the analysis of line emission and continuum maps is precluded, since it is a very complicated task to isolate the contribution of each component. Again, we refer to  \cite{Lindberg2014} for insight into the methods that can be used.

\begin{figure}[!h]
\centering
\includegraphics[]{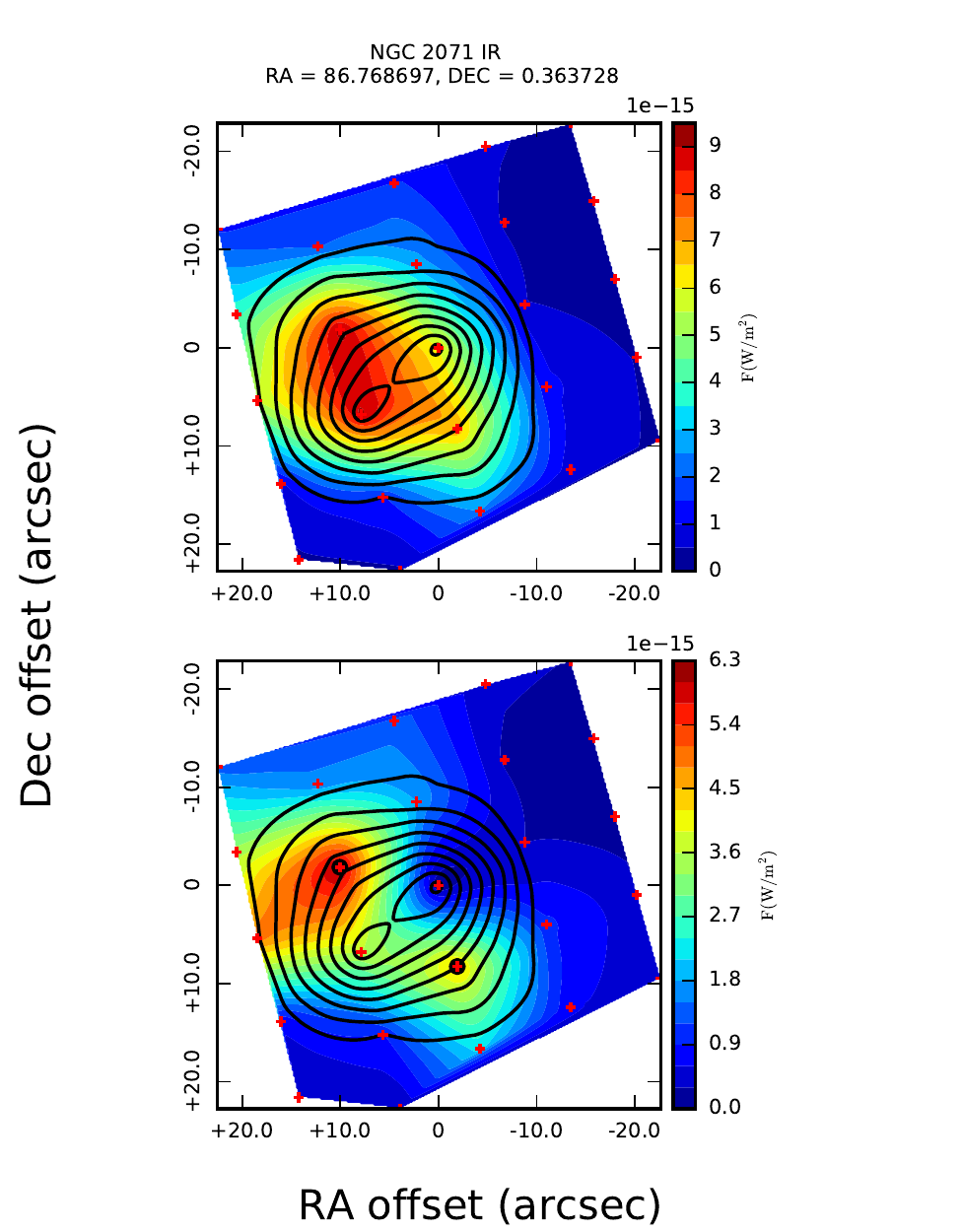}
\caption{Top: o-$\rm H_{2}O$ emission at 63 $\rm \mu m$ map for NGC 2071 IR.. Bottom: residual map for o-$\rm H_{2}O$ emission at 63 $\rm \mu m$ for NGC 2071 IR.  The solid line contours depict the continuum emission in both panels.}
\label{Fig:NGC2071_H2O_Maps}
\end{figure}

Only one source showed extended emission in \mbox{o-$\rm H_{2}O$} at 63.325 $\rm \mu m$, NGC 2071, a Class 0 source from the WISH program. Its line emission and residual maps are shown in Fig. \ref{Fig:NGC2071_H2O_Maps}. The source shows both [OI] and o-$\rm H_{2}O$ residual emission. [OI] residual emission is found south and north-east of the source, while o-$\rm H_{2}O$ residual emission is found southward only. Interestingly, [OI] residual emission is brighter in the north-east position than in the south position, and the maximum is $\rm \sim$ 10 times brighter than that of the o-$\rm H_{2}O$ residual. \cite{Melnick2008} detected extended warm $\rm H_{2}O$ emission in the region aligned with the direction of the outflow.

\begin{figure*}[!t]
\begin{center}
\includegraphics[]{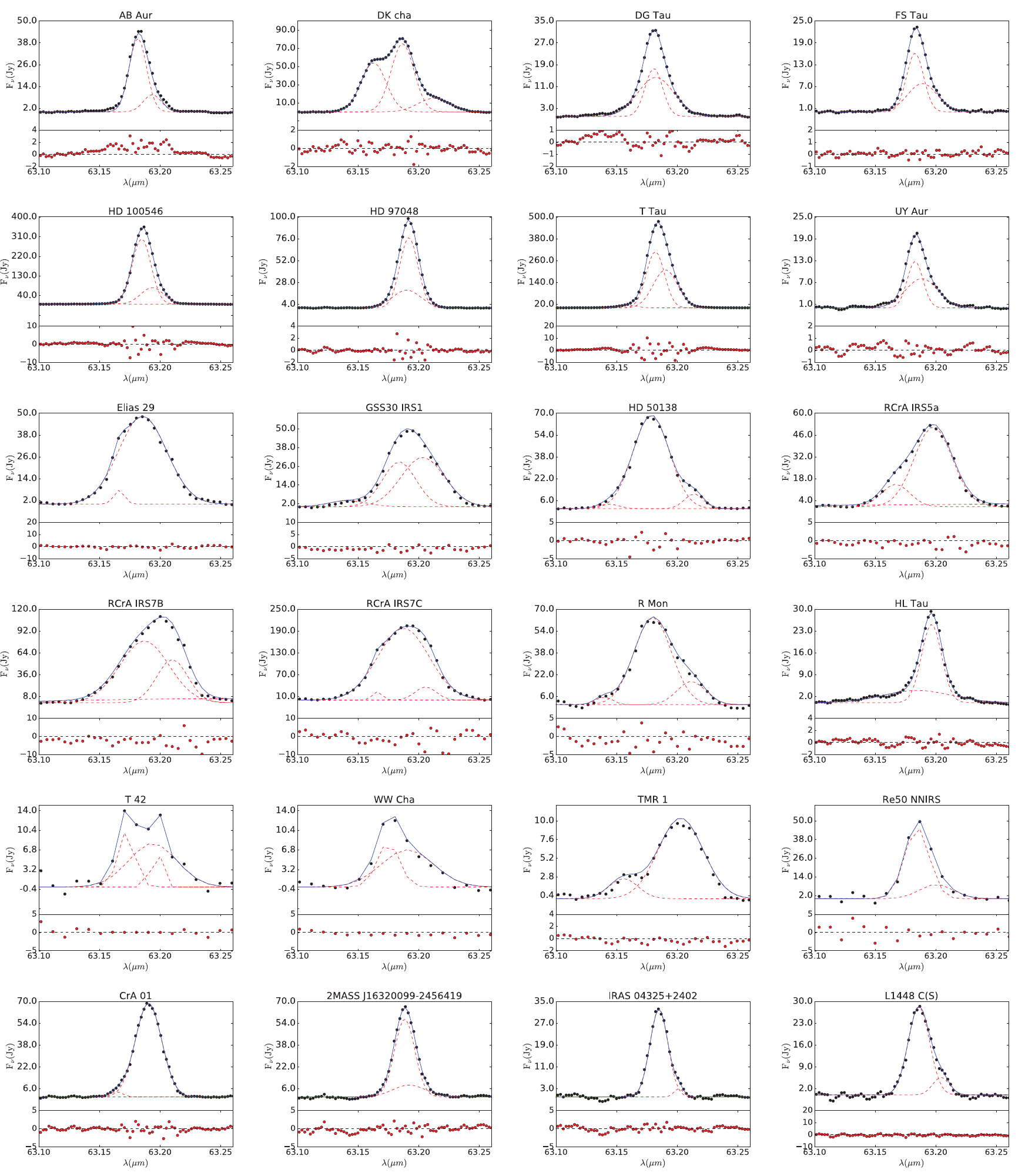}
\caption{Multiple-component fits for sources that are better fitted by a combination of two to three Gaussians. The red dashed lines show the individual Gaussian components, while the blue solid lines depict the combined model. Black dots represent the observed spectra. }
\label{Fig:multiG}
\end{center}
\end{figure*}

\addtocounter{figure}{-1}
\begin{figure*}[!t]
\begin{center}
\includegraphics[]{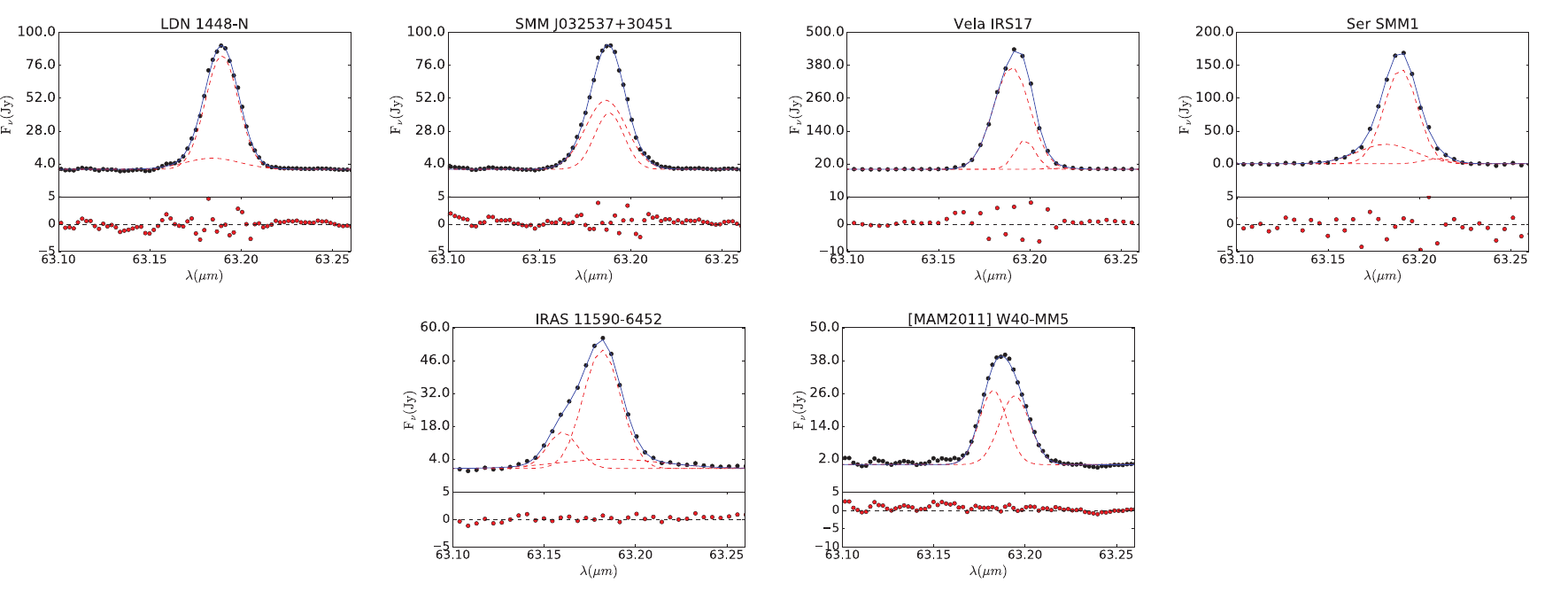}
\caption{continued.}
\label{Fig:multiG1}
\end{center}
\end{figure*}

\subsection{Multiple components}\label{Sec:MultiComp}
\cite{vanKempen2010} showed that the profile of the [OI] line at 63 $\rm \mu m$ towards HH 46 observed with PACS  consisted of three components: blue- and red-shifted components and a rest-frame velocity component. The authors also showed that the velocities of the red- and blue-shifted components are consistent with observations of jets in the near-IR and in the optical. In \cite{Riviere2015}, the [OI] emission line was fitted by multiple Gaussians in DK Cha, indicating that several dynamical components might be present. The different velocities can be attributed to different components. While the rest-frame emission seems to be associated with  an envelope and/or disc plus wind emission, the most natural explanation for high-velocity components is jet emission. 

We performed a multiple Gaussian analysis of the YSOs in the sample. Model  fits with one, two, or three Gaussian components were compared using  the Bayesian information criterion \citep[BIC, see][]{Feigelson2012} to detect the most representative one. The BIC is described for each model as
\begin{equation}\label{Eq:BIC}
\rm{BIC} = -2\times \ln(L^{0}(M)) + k\times \ln(N)
\end{equation}
where $L^{0}(M)$ is the highest value of the likelihood distribution, $k$ is the number of free parameters, and $N$ the number of spectral points. To detect the best model in each case, we performed a $\rm \chi^{2}$ minimization, and therefore Eq. \ref{Eq:BIC}  becomes

\begin{equation}\label{Eq:BIC2}
\rm{BIC} = \chi^{2}_{0} + k\times \ln(N)
\end{equation}
where $\chi^{2}_{0}$ is the lowest value of $\rm \chi^{2}$, corresponding to the highest value of the likelihood distribution. To decide which model better reproduces the observations, the BIC for one-, two-, and three-Gaussians models must be compared. A BIC difference $\rm 2< \Delta BIC<6$ shows evidence against the model with the higher BIC, while a BIC difference larger than 10 excludes the model with the higher BIC with high probability.

Thirty sources showed evidence of multiple components. We show in Fig. \ref{Fig:multiG} the resulting fits for these sources, and the Gaussian parameters are given in Table \ref{Tab:multiG}. It is a known effect that when a source is not properly centred on a PACS spaxel it can result in a shift in the line centre and in a distortion of the Gaussian shape. However, 24 of the 30 sources are properly centered on their spaxels, and therefore, at least for them, we are sure that the effect is real and not an observational artefact. The number of sources per evolutionary stage is as follows: one is a highly embedded source with unknown evolutionary stage, four are Class 0 sources, one is an intermediate Class 0/I sources, 12 are Class I, one is an intermediate Class I/II, 10 are Class II (six are T Tauri and four are HAeBe stars), and one is a transitional disc.  

Fourteen of them have a smaller separation between the different components than the spectral resolution at 63 $\rm \mu m$, $\rm \sim 88 ~ km/s$. Twenty sources are better reproduced by a model with two Gaussians, while twelve sources are better reproduced by a model with three components. The most prominent case is that of DK Cha \citep[see][]{Riviere2015}, where the separation between the different components in the central spaxel is evident.  Of the 30 sources that need multiple Gaussians to be fitted, 21 show evidence of extended emission according to their 3$\rm \times$3 fluxes and 5$\sigma$ residual emission in their IFUs.

\begin{figure}[!t]
\begin{center}
 \includegraphics[]{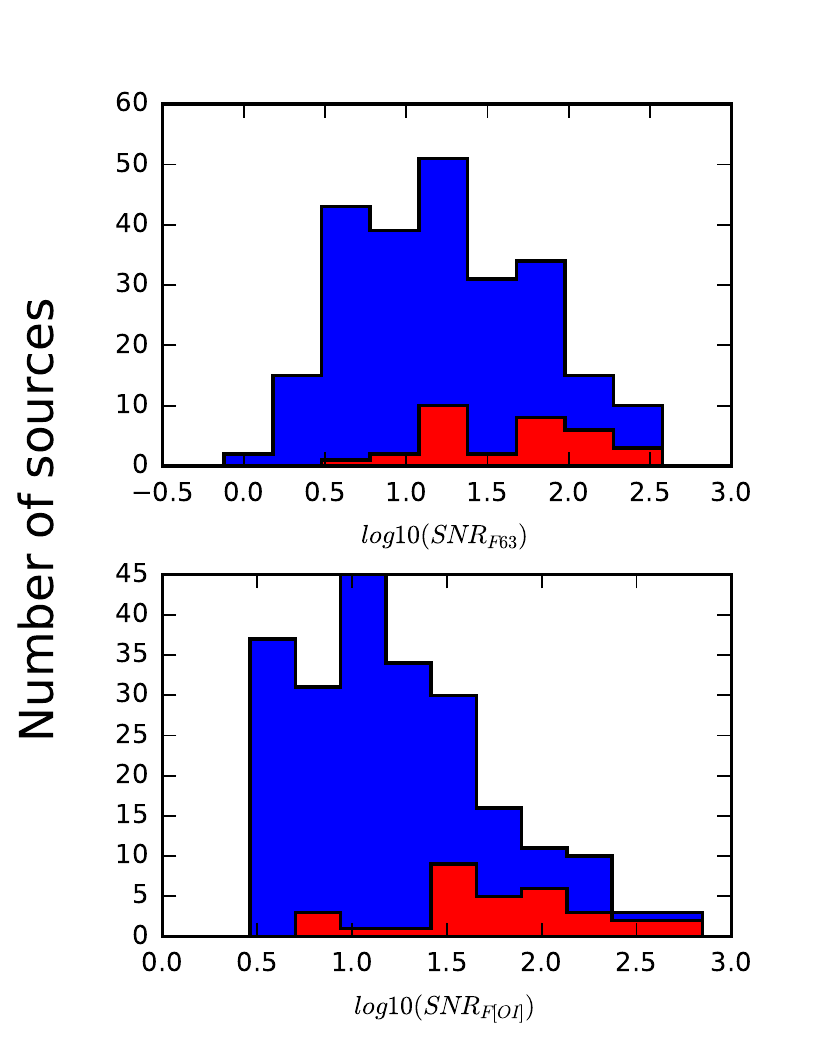}
\caption{Top: distribution of continuum S/N for the whole sample (blue) and for sources that are better reproduced by multiple Gaussians (red). Bottom: distribution of line S/N for the whole sample (blue) and for sources that are better reproduced by multiple Gaussians (red).}
\label{Fig:hist_S/N}
\end{center}
\end{figure}

To test wether detecting multiple components was linked to high S/N observations, we compared the distributions of continuum and line S/Ns of the whole sample with that of sources that are better reproduced by multiple Gaussians. Histograms comparing the distributions are shown in Fig. \ref{Fig:hist_S/N}, demonstrating that multiple-Gaussian detections are linked to high S/N sources. A Kolmogorov-Smirnov test confirmed this trend: the probability that both distributions are drawn from the same sample are $\rm \sim 10^{-3}$ for the continuum S/N and $\rm \sim 10^{-7}$ for the line S/N. Therefore, we cannot rule out that some, if not all, of the low-S/N sources  also have multiple components. 

\begin{figure*}[!t]
\begin{center}
\includegraphics[scale=0.25]{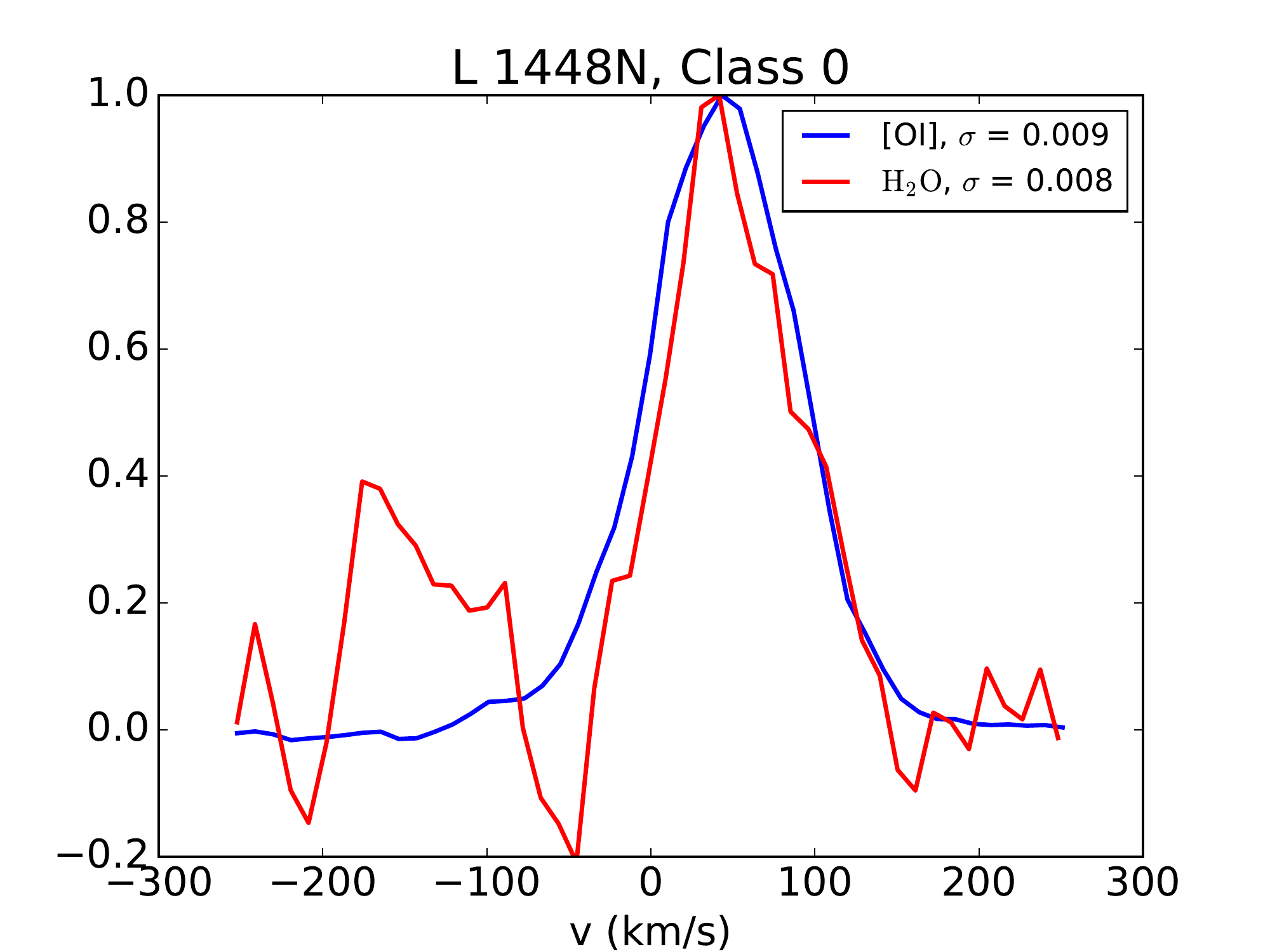}\includegraphics[scale=0.25]{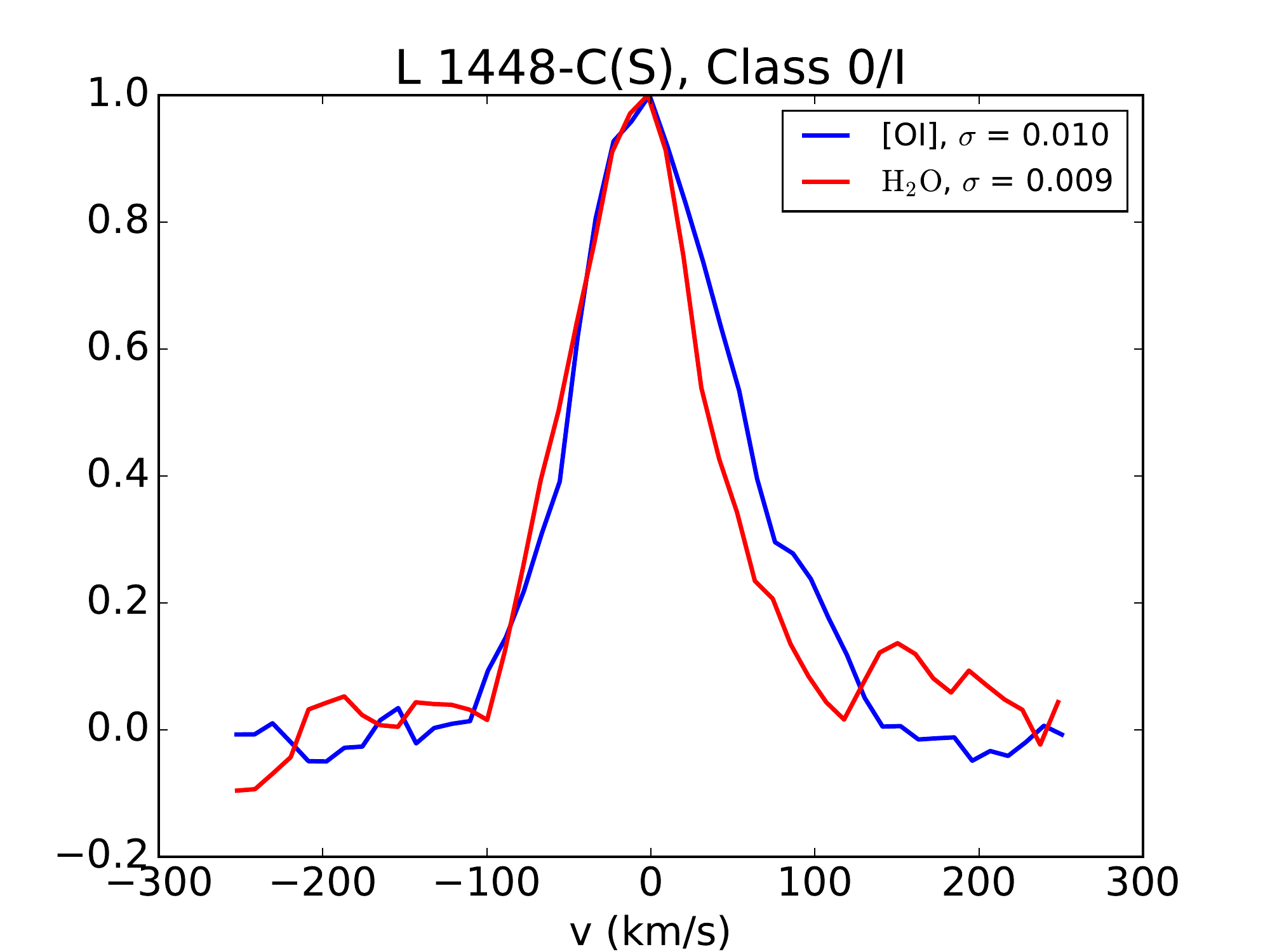}\includegraphics[scale=0.25]{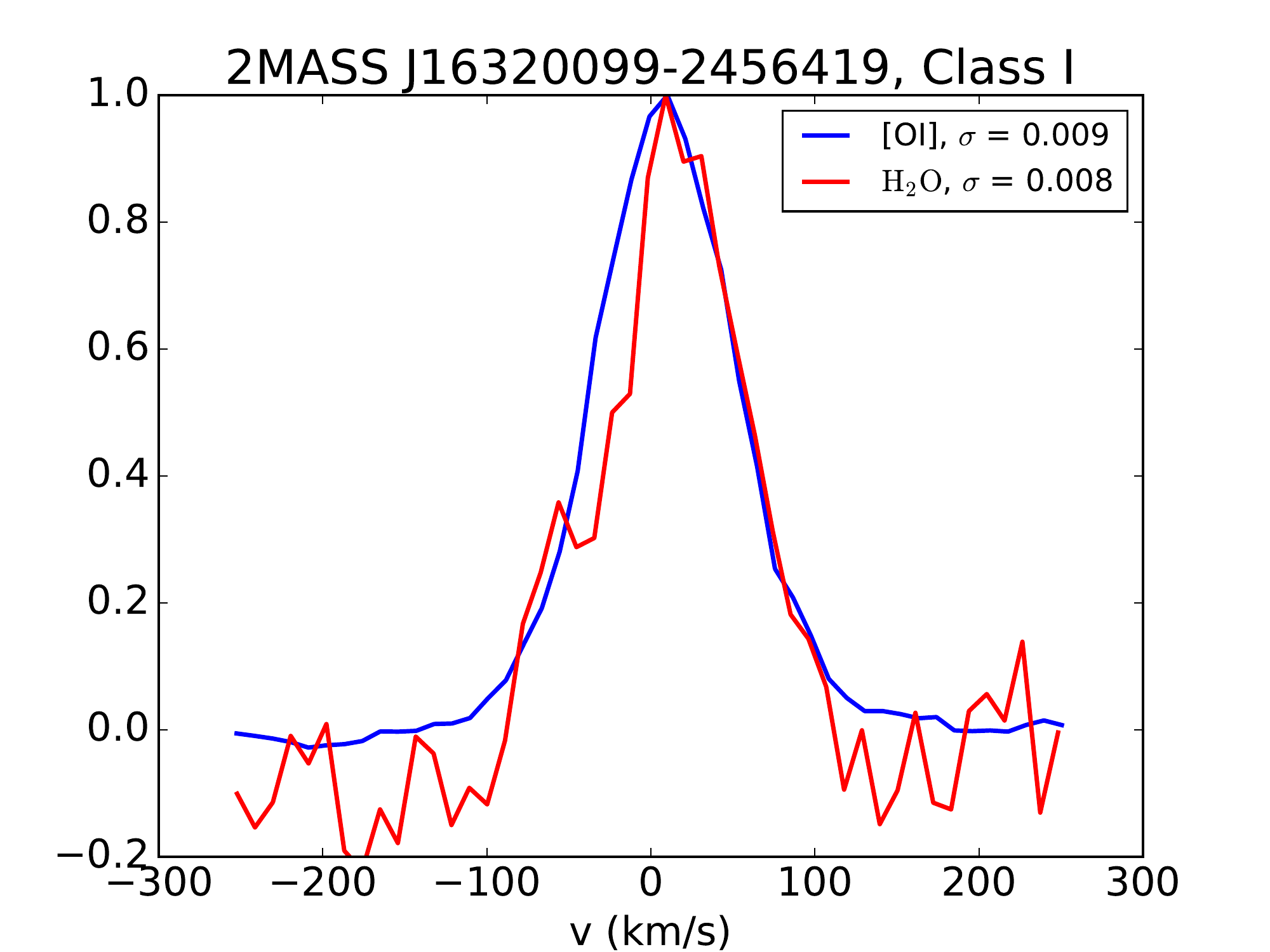}\\
\includegraphics[scale=0.25]{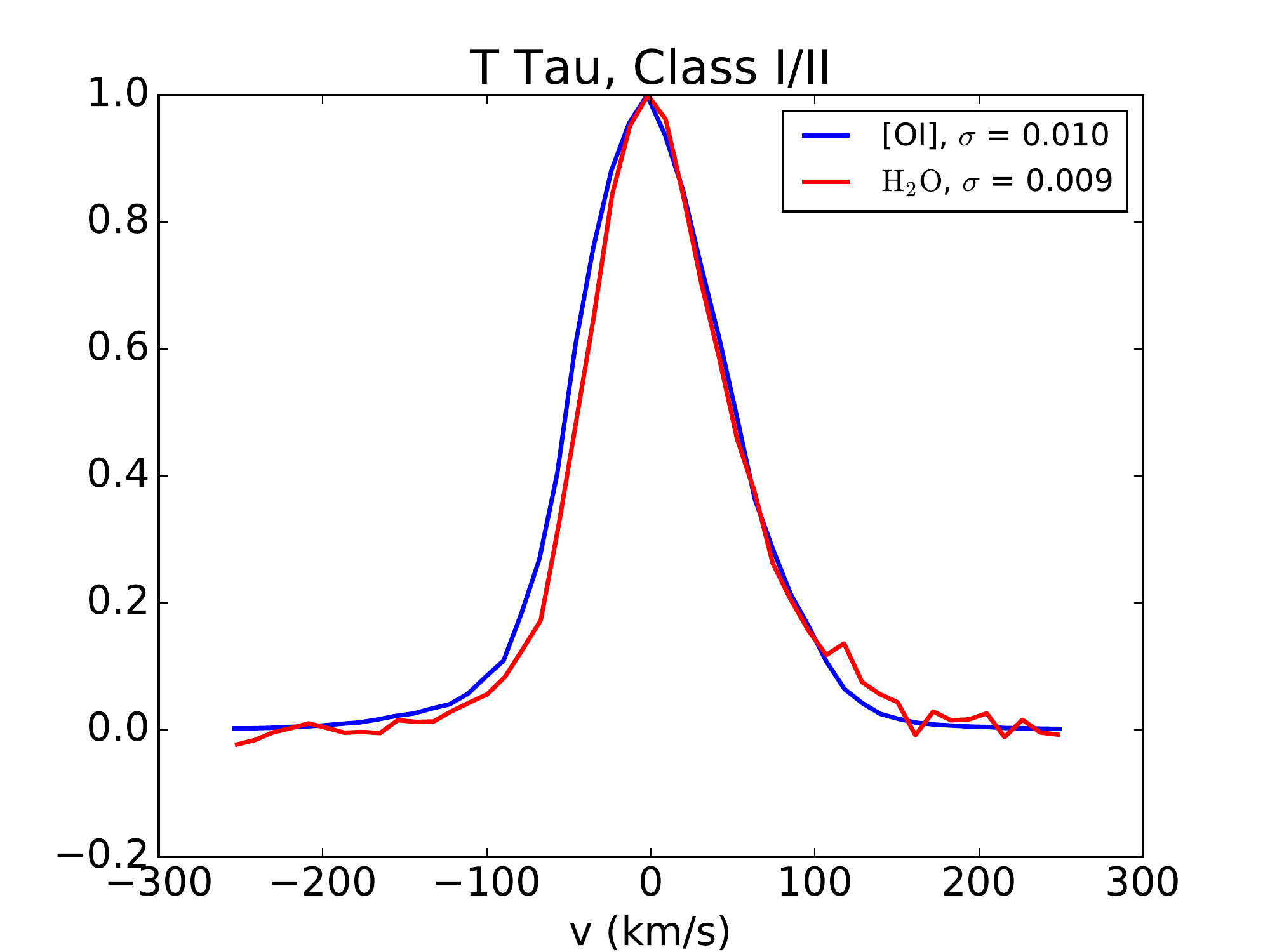}\includegraphics[scale=0.25]{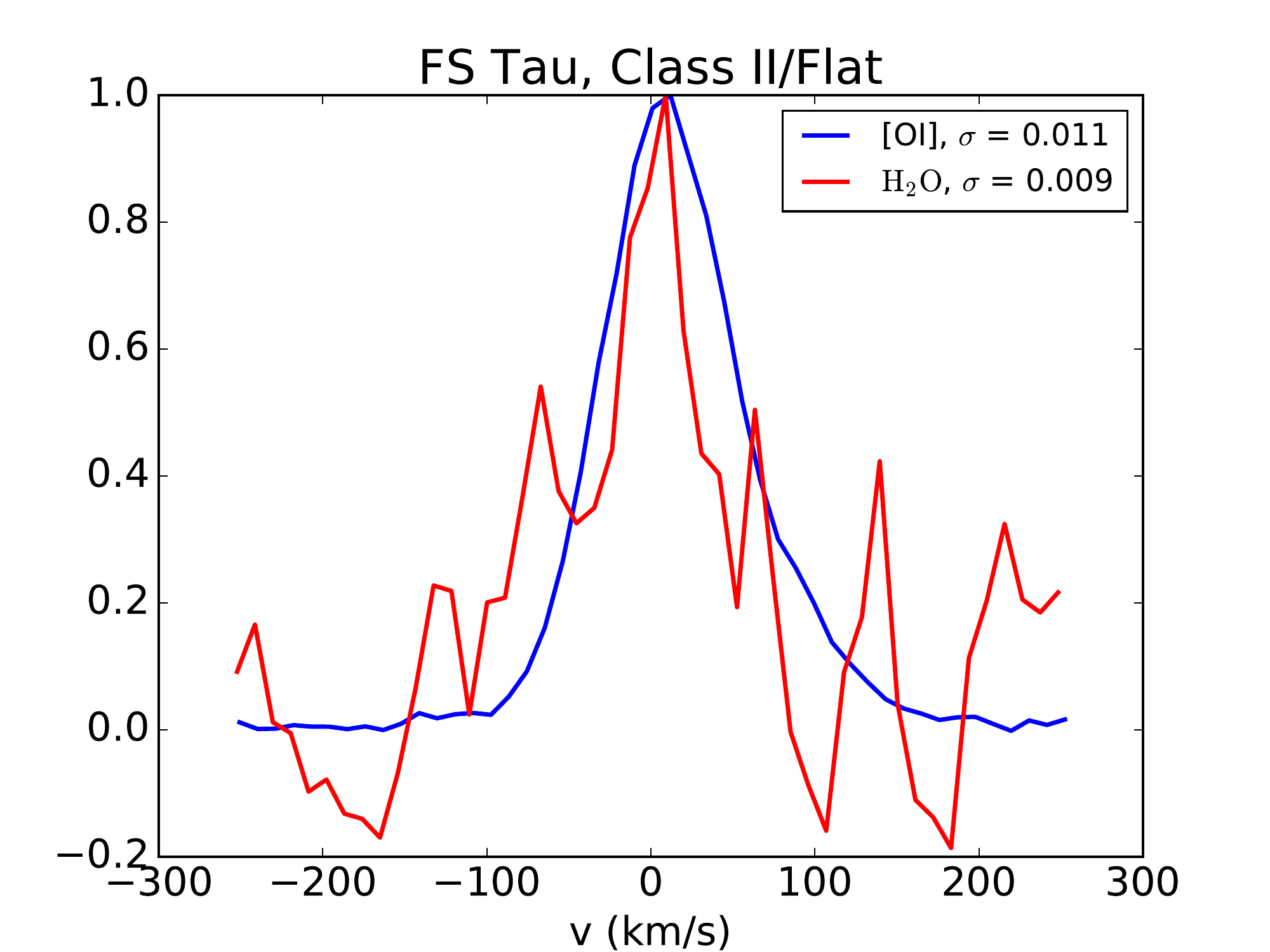}\includegraphics[scale=0.25]{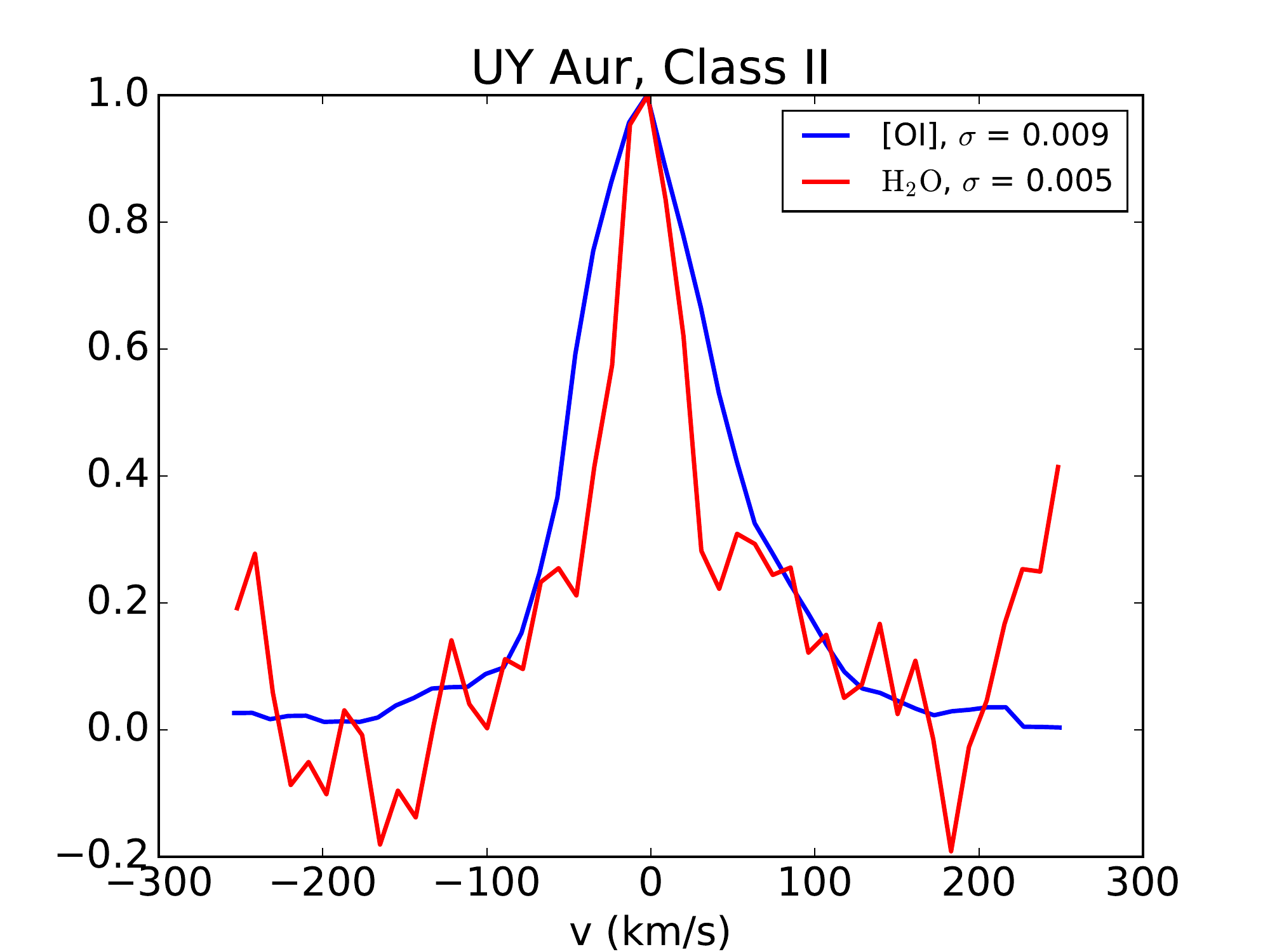}\\
\caption{Comparison of the [OI] and o-$\rm H_{2}O$ line profiles for sources whose [OI] profiles are better reproduced by multiple Gaussians. In the legends we show the measured $\sigma$ of the Gaussian fits.}
\label{Fig:CompSpProf}
\end{center}
\end{figure*}

\section{Discussion}
As shown by \cite{Podio2012}, [OI] emission can be extended along the jet direction, indicating that the jet contributes to, or even dominates, the line emission. We computed in Sect. \ref{Sec:ExtEmTest} residual emission maps for sources in the sample, and showed that [OI] is extended for a large number of sources: 83 sources show extended emission when the flux from the central spaxel is compared to that of co-added ones, and 71 of them showed residual emission. We also showed that $\rm H_{2}O$ emission is extended in only one case, compared to [OI], an indication that the o-$\rm H_{2}O$ line most likely has its origin in a compact and high-density region at the base of the jet \citep[see][]{Podio2012}, like the disc or shocks along the cavity walls in the envelope. Aiming to better understand the physics behind the line emission at 63 $\rm \mu m$ and its origin, we compared line fluxes and different observables. To minimize the scatter associated with the different distances, all the fluxes where scaled to the distance to Taurus \citep[140 pc, ][]{Kenyon2008}, so that only sources belonging to known associations or with known distances were used. The distances for the different associations are given in Table \ref{Tab:assoc}.

\begin{table}[!t]
\caption{Distances to associations in the sample}             
\label{Tab:assoc}      
\centering          
\begin{tabular}{lcc}     
\hline \hline 
Association & Distance & Reference \\
--  & (pc) & -- \\
\hline
BPMG & 33 & \cite{Zuckerman2004}$^{*}$ \\
Cha & 165 & \cite{Luhman2008} \\
Cha II & 178 & \cite{Luhman2008} \\
CrA & 130 & \cite{Neuhauser2008}\\
$\eta$ Cha & 97 & \cite{Mamajek1999} \\
Lupus & 150 & \cite{Comeron2008} \\
Lupus III & 200 & \cite{Comeron2008} \\
Oph &130 & \cite{Wilking2008} \\
Per & 235 & \cite{Hirota2008}\\
Serpens & 415 & \cite{Dzib2010} \\
Taurus & 140 & \cite{Kenyon2008} \\
TWA & 50 & \cite{Webb1999} \\
Tuc Hor & 46 & \cite{Zuckerman2004}$^{*}$ \\ 
Up Sco & 145 & \cite{Zeeuw1999}\\
\hline                  
\end{tabular}
\tablefoot{(*): the distance provided is the average of the distances to individual sources shown in the referenced papers.}
\end{table}

\subsection{Correlation between far-IR line emission fluxes}
\begin{figure}[!t]
\begin{center}
\includegraphics[]{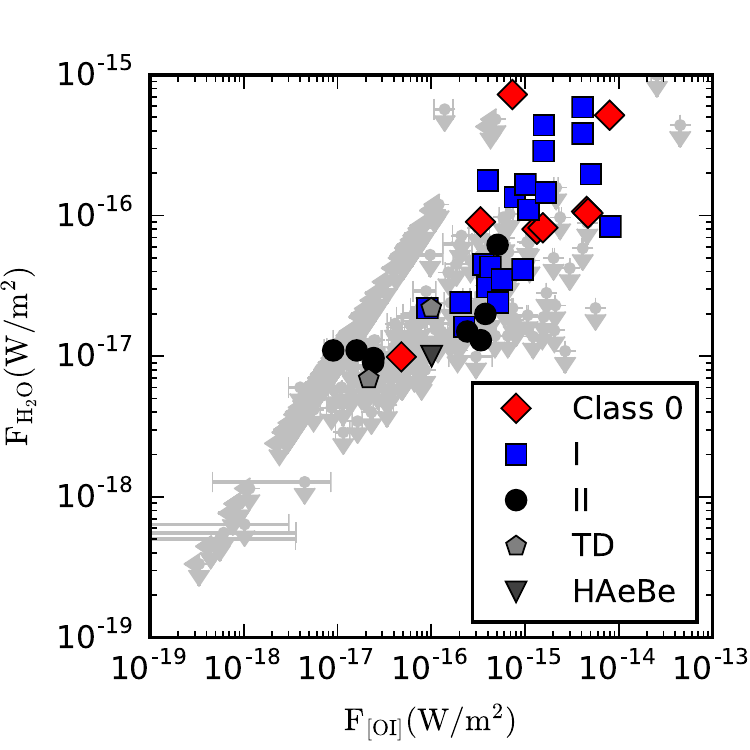}\\
   \caption{o-$\rm H_{2}O$ line flux versus [OI] line flux at 63 $\rm \mu m$. Arrows show the positions of upper limits for non-detected sources. All fluxes have been scaled to the distance of Taurus.}
   \label{Fig:H2O_vs_OI}
\end{center}
\end{figure}

In Fig. \ref{Fig:H2O_vs_OI}, we show the relation between o-$\rm H_{2}O$ line flux and [OI] line flux at 63 $\rm \mu m$. This correlation was previously found by \cite{Riviere2012}. With better source statistics, we now tentatively see a change in slope from Class 0 and I to Class II sources. However, the small number of o-$\rm H_{2}O$ detections precludes any firm conclusion. The fact that [OI] and o-$\rm H_{2}O$ fluxes at 63 $\rm \mu m$ are correlated might suggest a common origin for the two lines lines. However, while [OI] is sometimes extended, the o-$\rm H_{2}O$ line is extended in only one source. Furthermore, while we sometimes need multiple components to fit the [OI] line, a simple Gaussian fit is enough to fit the o-$\rm H_{2}O$ line in all detections. However, the lack of multiple components can be linked to the low-S/N nature of the detections. Adding more evidence against a co-spatial origin, the critical density for the o-$\rm H_{2}O$ line is orders of magnitude higher than the one for [OI]. 

The lack of multiple-components in the profile of o-$\rm H_{2}O$ might be due to low S/N. o-$\rm H_{2}O$ detections with a line flux S/N similar to that of [OI] lines that require multiple Gaussians (such as L1448-C(S), $\rm S/N_{fH_{2}O}\sim24$), are well reproduced by a single Gaussian. We show in Fig. \ref{Fig:CompSpProf} a comparison of the [OI] and $\rm H_{2}O$ line profiles for sources whose [OI] profiles are better reproduced by a combination of Gaussians. L1448-C(S) and 2MASS J16320099-2456419 show similar [OI] and $\rm H_{2}O$ profiles, with bumps at similar velocities ($\rm \sim50-100 ~km/s$), but narrower $\rm H_{2}O$ lines. By contrast, L 1448N, FS Tau and UY Aur show very different profiles. T Tau show similar shapes for both lines, but the $\rm H_{2}O$ shows a narrower profile and no bumps are seen. Observations with high spectral resolution are needed to  explain multiple components.

The most likely explanation for the spatial extension of the emission, and for the presence of multiple components in the [OI] line is a contribution from jet emission associated with the source \citep[][]{Podio2012,Howard2013}. The high-velocity components ($\rm v_{HVC} \sim 100 km/s$) must be associated with jets, while the low-velocity ones (which might also be rest-frame velocity, due to the limited spectral resolution) can be associated with envelope and wind emission, as well as disc emission.

o-$\rm H_{2}O$ seems to be dominated by disc or envelope emission or compact jet emission, since  extended emission is observed in only one Class I source, and the line profiles are consistent single Gaussians. However, we mostly detected water emission in jet sources, and therefore we cannot rule out that the emission originates at the base of the jet, and that the lack of multiple components is due to a low S/N in the observations presented here.

\subsection{Correlations with continuum emission}\label{Subsec:correlationsWithCont}
\begin{figure}[!t]
\begin{center}
 \includegraphics[]{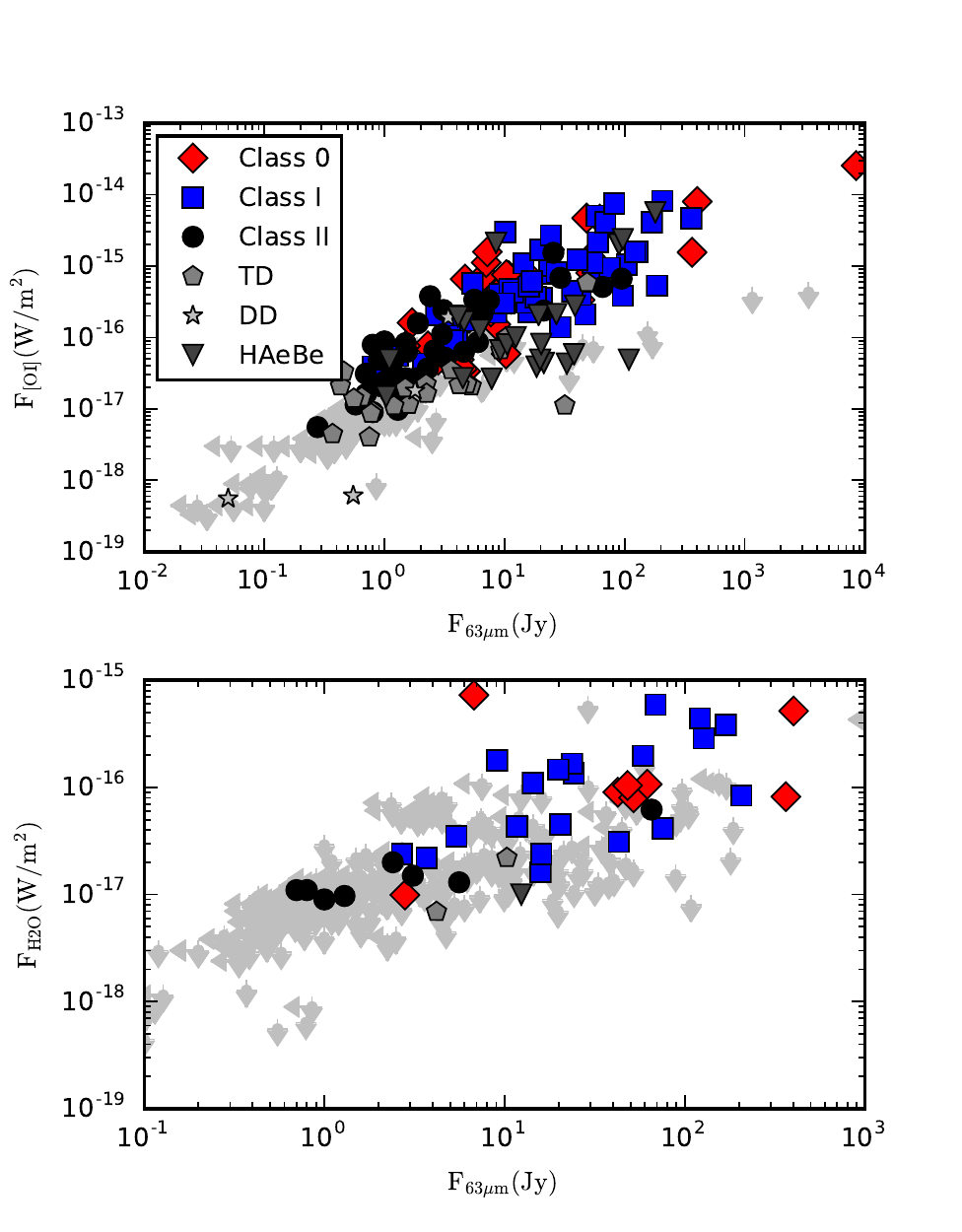}
   \caption{[OI] (top) and o-$\rm H_{2}O$ (bottom) line fluxes versus continuum flux at 63 $\rm \mu m$ for sources in the sample. Sources are labelled according to evolutionary stage. All the fluxes have been scaled to the distance to Taurus (140 pc). Arrows show the position of upper limits for non-detected sources. }
 \label{Fig:fOI_vs_cont}
\end{center}
\end{figure}

\cite{Howard2013} studied [OI] emission in Taurus and found a correlation between [OI] line emission at 63 $\rm \mu m$ and the continuum at the same wavelength. This correlation was later confirmed for other associations \citep{Mathews2013,Riviere2015}.  We have extended the study to the entire sample of YSOs observed with PACS in spectroscopic mode. The resulting plot is shown in Fig. \ref{Fig:fOI_vs_cont}. The different sources are placed at different loci in the diagram. Class 0 and I sources show on average higher line fluxes for the same continuum level than Class II and transitional discs. Furthermore, transitional discs show also lower [OI] fluxes for the same level of continuum emission than Class II sources. Transitional discs are located in the lowest part of the diagram because theirs is the lowest [OI] emission for the same continuum level. Class II sources show intermediate [OI] fluxes, while the Class 0 and I sources show the highest flux levels for the same continuum. 

\begin{figure*}[!t]
\begin{center}
\includegraphics[]{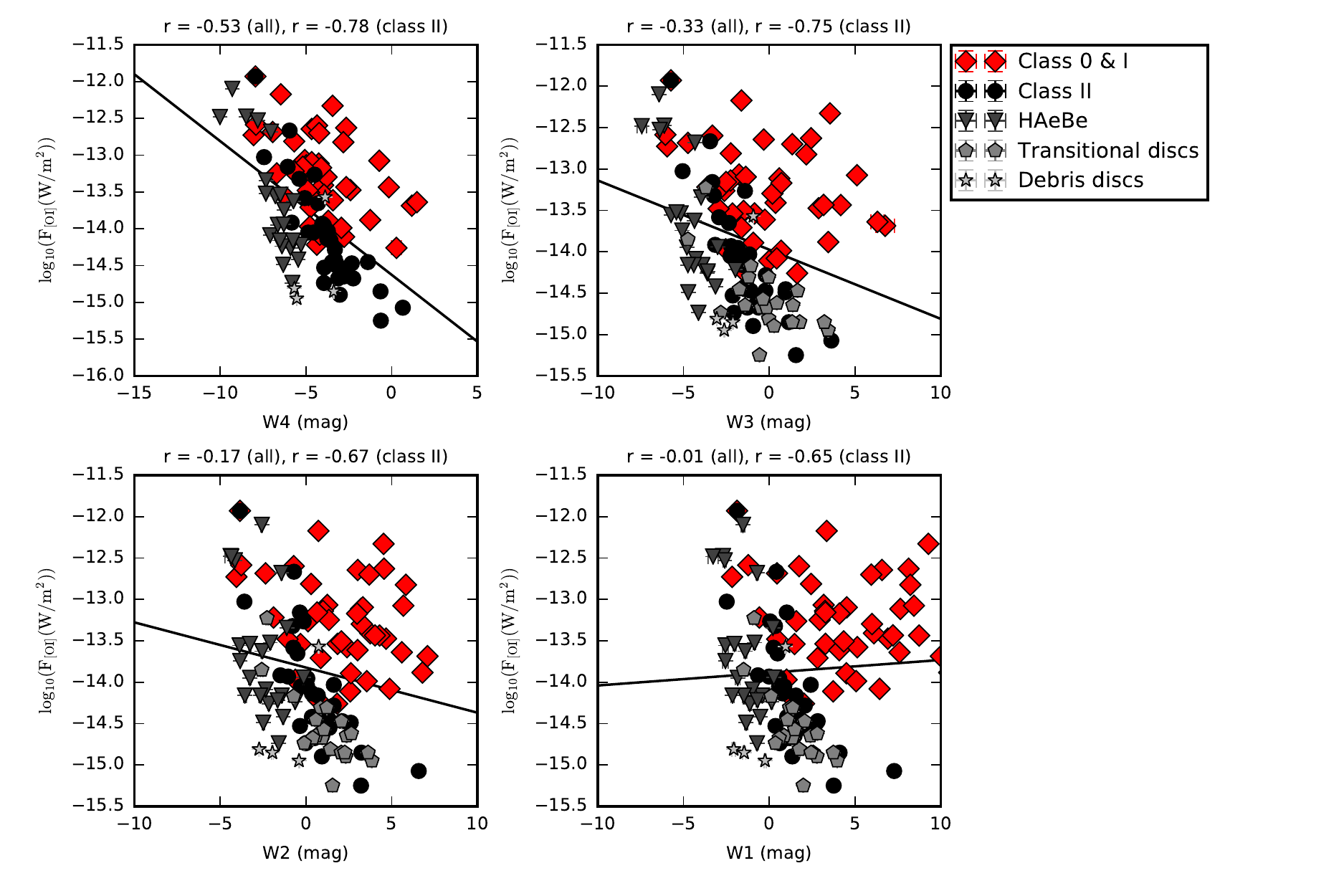}
   \caption{[OI] fluxes at 63 $\rm \mu m$ versus WISE magnitudes. All the fluxes have been scaled to the distance to Taurus. Magnitudes are corrected for distance. Spearman correlation coefficients are shown at the top of each plot.}
   \label{Fig:fOI_vs_WISE}
\end{center}
\end{figure*}

The correlation between [OI] flux and the continuum flux at 63 $\rm \mu m$ is different for jet and non-jet sources \citep{Howard2013}. Given the correlation between line emission and continuum emission at 63 $\rm \mu m$, we tested correlations against continuum photometry at different wavelengths. We first started by retrieving WISE  \citep{Wright2010} magnitudes for sources in the sample, within a search radius of 2.6\arcsec. When testing the correlations, we only included sources belonging to associations with known distances, or belonging to known associations, to correct the fluxes and magnitudes for distance. All the fluxes were scaled to the distance to Taurus (140 pc), and the magnitudes were converted into absolute magnitudes. The resulting comparison is shown in Fig. \ref{Fig:fOI_vs_WISE}. We observed clear correlations between the [OI] flux and WISE band 4 flux, but the correlation seems to vanish for WISE1, with a Spearman probability for the null hypothesis (i. e. that there is no correlation) that decreases from 3.4 $\rm \mu m$ (WISE 1) to 22 $\rm \mu m$  (WISE 4). However, the results change when only Class II sources are considered, and the correlation is present for all the WISE bands, with strong to very strong correlation coefficients in the range 0.7 to 0.8.

The scatter in the correlations has many contributions, such as instrumental uncertainties, scatter in distance within the same association, and different disc mass and geometry. Furthermore, jets and winds also contribute to the [OI] line flux, but not to the continuum, which increases the scatter. We observed clear correlations from 22 to 63 $\rm \mu m$. The fact that [OI] correlates with magnitudes at different IR wavelengths, and that the strength of the correlation increases with wavelength, and is more pronounced for Class II sources very likely indicates that dust at different temperatures and gas emission are related, which most likely points to a contribution from discs and envelopes. However, this is not the only interpretation. Sources accreting at higher rates will show brighter continuum emission at 63 $\rm \mu m$. If accretion is driving [OI] emission at 63 $\rm \mu m$, then sources with higher accretion rates will also show brighter [OI] emission at 63 $\rm \mu m$, explaining the correlation.

\subsection{Comparison with DENT models}\label{Sec:discModels}
To better understand the correlations, we compared our results with predictions from the DENT grid \citep{Woitke2010,Kamp2011}. The DENT grid consists of more than 3$\times 10^{5}$ models of protoplanetary discs that were developed to help in the interpretation of photometric and spectroscopic observations of protoplanetary discs for the GASPS program \citep{Dent2013}. The grid contains models representing different evolutionary stages. To restrict the number of models and interpret the results, we fixed $\rm R_{in}=R_{sub}$ (where $\rm R_{sub}$ is the dust sublimation radius), $\rm R_{out}=300~au$, a surface mass density distribution power-law with index $\rm \epsilon_{0}=1.0$, the gas-to-dust ratio to 100 and the minimum grain size to 0.05 $\rm \mu m$. We also excluded edge-on models ($i$ = 90$\rm ^{\circ}$) and models with $\rm \beta=0.8$ (where $\beta$ is the exponent of the scale height relation $H = H_{0}(r/r_{0})^{\beta}$, leaving us with $\rm \beta = 1.0$ and $\rm \beta = 1.2$).

\begin{figure}[!t]
\begin{center}
\includegraphics[]{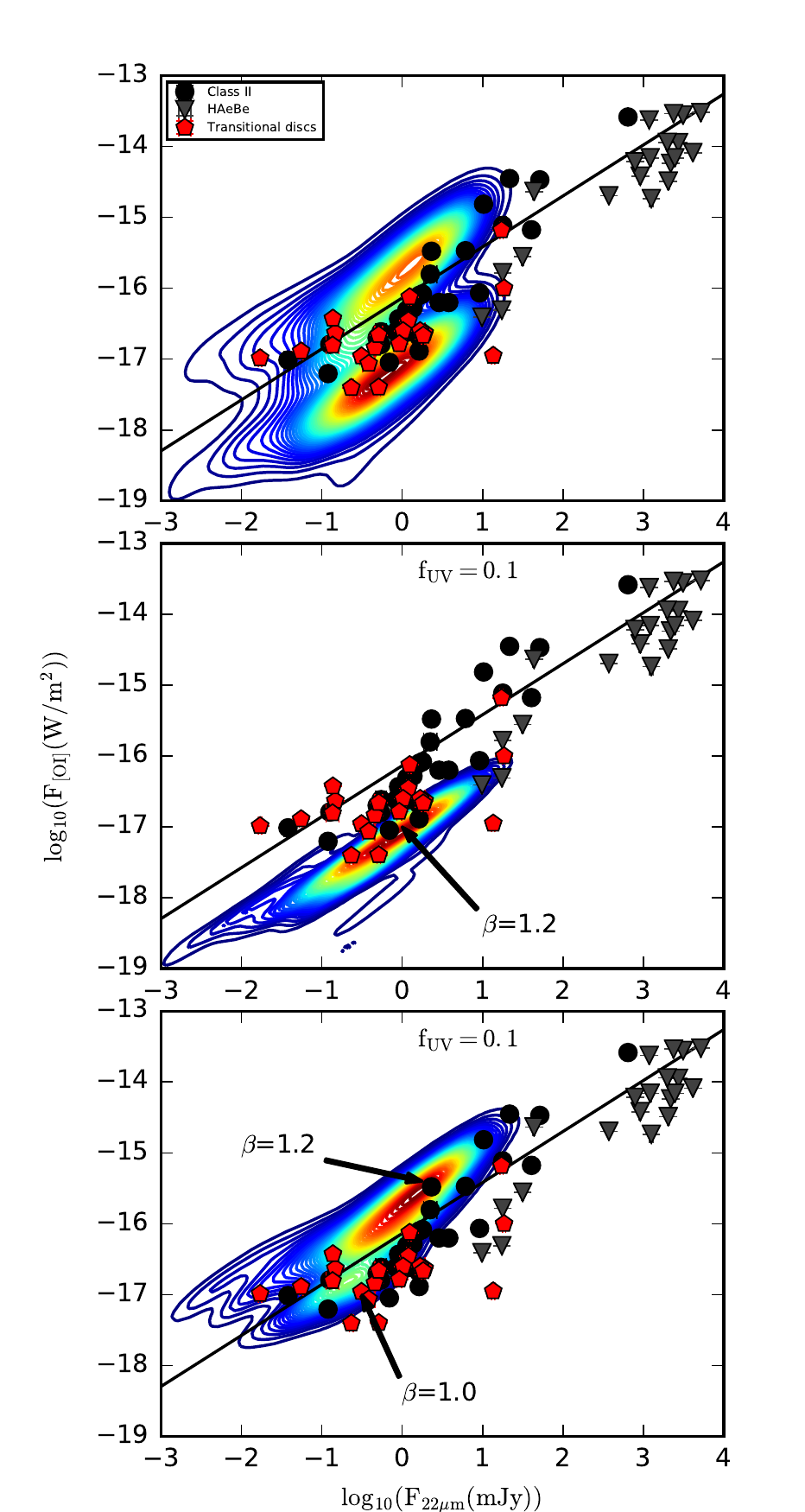}
   \caption{[OI] fluxes at 63 $\rm \mu m$ versus WISE flux at 22 $\rm \mu m$: observations compared to models. The contours show the density of points for models from the DENT grid. Black dots show the position of observed Class II sources, red pentagons are transitional discs, and grey triangles are HAeBe stars. The top panel shows the whole distribution of selected models. The middle panel shows the distribution of models with $\rm f_{UV}=0.1$, while the bottom panel shows models with  $\rm f_{UV}=0.001$. The arrows point to the location of sources with different values of $\beta$. The solid line is a linear fit to the observed data.}
   \label{Fig:fOI_vs_WISE_models}
\end{center}
\end{figure}

The DENT grid does not include continuum fluxes at 22 $\rm \mu m$, but at 24 $\rm \mu m$, so that this is what we show in Fig. \ref{Fig:fOI_vs_WISE_models}. The difference in wavelength is so short that any effect on the shape of the correlation must be small. Owing to the limited parameter space covered by the DENT grid, the brightest sources are not covered by the models. The spearman probability (p) for the null hypothesis (i. e., that there is no correlation) is  $p \ll $$\rm1.0^{-3}$ for the observations and for the models. We then fitted a straight line in the log-log space to the distributions of models and observations. The resulting fit has slope $m$=0.72 for the observational distribution, and $m$=0.61 for the distribution of models. 

According to \cite{Woitke2010} the UV radiation field is one of the main drivers of [OI] emission, together with the flaring geometry and the total gas mass. To better understand the influence of these parameters on the [OI] flux distribution, we performed a more detailed analysis by fixing one of them at a time, and letting the other parameter free. The resulting comparison is shown in Fig. \ref{Fig:fOI_vs_WISE_models}. For  $\rm f_{UV}=0.001$, only models with $\rm \beta=1.2$ can reproduce the observations. When $\rm f_{UV}=0.1$, models with  $\rm \beta=1.2$ overestimate the flux for most of the observations. It is clear from the plots that we need intermediate values of both $\rm f_{UV}$ and $\beta$. However, extremely flared discs with low $\rm f_{UV}$ or flat discs with very high $\rm f_{UV}$ overlap with the observations. Overall, DENT models provide a good description of [OI] emission at 63 $\rm \mu m$ for Class II sources. This shows that observations of Class II sources are compatible with pure protoplanetary disc models. Furthermore, the DENT models do not include jets or outflows, but they describe the emission well. The theoretical prediction by \cite{Woitke2010} that in a disc $\rm f_{UV}$ and $\beta$ control [OI] emission is therefore compatible with our observations. We conclude that disc models provide an explanation for the correlations, but other solutions cannot be excluded.

\subsection{Origin of line emission}
In a study of molecular and atomic emission towards \object{HH~46}, \cite{vanKempen2010} showed that the bulk of [OI]  emission comes from low-velocity gas, after the impact of high-velocity jets in the cavity walls. The authors also concluded that the high-velocity component observed in the central and outer spaxels originates in fast dissociative shocks in the lower density jet. \cite{Podio2012} demonstrated that atomic [OI] and [CII] line emission were extended and correlated with the direction of optical jets, and proposed that the extended atomic emission could be produced by J-shocks. \cite{Karska2013} studied a sample of low-mass Class 0 and I YSO and concluded that [OI] emission at 63 $\rm \mu m$ originates in dissociative shocks. The authors also distinguished two groups of sources with extended emission, based on morphological differences: a compact group, where [OI] and OH emission dominates the central spaxel, while CO and $\rm H_{2}O$ can follow the same trend or are dominated by off-source emission, and an extended group, where OH off-source emission is strong. Class 0 sources dominate the extended group while Class 0 and I are equally represented in the compact group. The ratios of [OI] line emission at 63 and 145 $\rm \mu m$ computed by \cite{Lee2014} for six low-mass embedded sources in Taurus were consistent with an origin in C-shocks, and again the authors highlighted that atomic emission is commonly extended along the jet direction. Additional support for a jet origin comes from \cite{Nisini2015}, where the very similar profiles shown by [OI]  and [FeII] towards LDN 1448N were highlighted. The authors were able to separate the contributions from the different dynamical components in a few favourable cases, and observed an increment in [OI] velocity with distance from the central source, in agreement with observations of SiO.

In the present paper we have studied the spatial distribution of [OI] line emission towards 110 Class 0 and I sources, and detected hints of extended emission in at least 60 of them ($\rm 55\% \pm 5$). When separated by groups (see Fig. \ref{Fig:histExt}), we found a slightly larger extended emission fraction for Class 0 ($\rm 0.63^{+0.07}_{-0.08}$) than for Class I ($\rm 0.51\pm 0.07$), although they are compatible within the errors. The fraction then dramatically decreases for Class II sources ($\rm 0.17\pm 0.08$, where we included Class II, both HAeBe and T Tauri, and transitional discs). We interpret this as a clear evidence that extended emission in Class 0, I, and II sources is due to shocks along the jet direction, in agreement with the results discussed in the previous paragraph, and that outflow activity drops with age.

Additional evidence for a jet contribution comes from the fact that 30 sources needed multiple Gaussians to fit the line profile. The reason for the low number is that only the most favorable cases will result in broad complex profiles because of the limited spectral resolution of PACS. Therefore, it is clearly established that extended emission is due to a jet contribution, and that at least part of the compact emission is also due to shocks at the base of the jet, near the compact source.

However, various authors explained [OI] emission in Class II sources solely by means of disc emission \citep[see, e. g.][, among others]{Thi2010,Woitke2011,Meeus2010,Thi2013,Thi2014,Tilling2012}.  Early results by \cite{Mathews2010} ruled out an outflow shock origin for the [OI] line emission observed towards HD 169142 and showed that TWA 01 and RECX 15 observations required extreme outflowing fractions to explain [OI] emission through outflow activity. Furthermore, the authors highlight edthat there is no evidence of outflows in these sources, and the spectral profiles for RECX 15 and TWA 01 are not resolved, precluding the presence of jets with line-of-sight velocities higher than 45 $\rm km~s^{-1}$. \cite{Gorti2011} modelled in detail a large set of emission lines towards TWA 01, from UV to radio emission, and concluded that [OI] line emission comes from a region that covers almost the whole disc (30-120 au). The comparison of [OI] line emission versus continuum emission at 63 $\rm \mu m$ performed by \cite{Howard2013} for Taurus sources leads to different loci for jet sources and sources without known jets. Jet sources show fluxes up to 20 times larger than than the non-jet sources. Furthermore, the authors discussed that transitional discs show even fainter fluxes, a trend that was confirmed by  \cite{Keane2014}, who studied [OI] emission in 17 transitional discs and demonstrated that they show weaker emission than full Class II discs, tentatively attributing the difference to flatter and/or less massive transitional discs compared to full ones.

In Sect. \ref{Sec:discModels} we have shown that radiative transfer models of protoplanetary discs can explain [OI] emission in Class II sources, adding evidence for a disc contribution. \cite{Gorti2008} showed that [OI] emission can be emitted by the surface of protoplanetary discs at all radii. Protoplanetary disc modelling of HAeBe stars by \cite{Kamp2010} predicted [OI] emitting regions extending from 30 to 100 au. Correlations between line emission and the continuum emission from 4.6 to 63 $\rm \mu m$ shown in Sect. \ref{Subsec:correlationsWithCont} argue in favour of an extended emitting region, and for similar spatial origins for the line and continuum emission, additionally supporting the likeliness of a disc contribution. Transitional discs in Fig. \ref{Fig:fOI_vs_cont} lie in the lower envelope of the cloud of points. We consider that they probably represent the real disc contribution, since no strong jet activity is expected in these sources. 

Overall, there is strong observational evidence supporting contributions from both the jet (in Class 0, I and II sources) and the disc (Class II sources), while \cite{vanKempen2010} ruled out a contribution from the passively heated envelope present in Class 0 and I sources. 

\section{Summary and conclusions}\label{ref:SumConc}

We have compiled \textit{Herschel}-PACS observations of [OI] and o-$\rm H_{2}O$ at 63 $\rm \mu m$ in YSOs, including Class 0, I, II, transitional discs, and debris discs, for a total of 432 observations of 362 sources.

We note that the [OI] emission line intensity, as well as detection fractions, decreases during the evolution from Class 0 to debris discs. However, we did not see a difference in [OI] emission between Class 0 and Class I, nor between Class II and transition discs. o-$\rm H_{2}O$ emission line intensity also decreases from Class 0 and I to more evolved sources (Class II and transition discs).

By means of comparing the fluxes computed from the central spaxel, the central 3x3 spaxels and the integrated IFU, we detected extended emission in the [OI] line for a total of 77 sources. For those sources showing hints of extended [OI] emission, we obtained line emission maps and residual maps, and confirmed residual emission in 71 sources. The fraction of sources showing extended emission decreases dramatically from Class 0, where 63\% of the sources show extended emission, to Class II, where only 17\% of the sources show extended emission.

We detected extended o-$\rm H_{2}O$ line emission in only one source.

For 30 sources in the sample we were able to fit multiple components to the line emission profile, which is indicative of different contributions to the line (envelope, discs, winds, and jets).

We have tested previously identified correlations in the entire sample. The [OI] line emission correlates with continuum emission at 63 $\rm \mu m$ for all classes, with the exception of of debris discs. 

We confirm the correlation between [OI] and o-$\rm H_{2}O$ at 63 $\rm \mu m$, and tentatively see a change in slope in the correlation between class 0 and I sources and class II sources.

We have identified new correlations with continuum emission between 4.6 and 22 $\rm \mu m$, indicating an extended emitting region (from the inner disc to tens of au) as the origin of the disc contribution.

\onllongtab{1}{
\begin{longtable}{llllll}
\caption{\label{YSO_sample} Sample of YSOs observed with \textit{Herschel}-PACS}\\
\hline\hline
Source name & RA & Dec & Sp. type$^{1}$ & Disc type & Association \\
-- & (deg) & (deg) & -- & -- & -- \\
\hline
\endfirsthead
\caption{continued.}\\
\hline\hline
Source name & RA & Dec & Sp. type  & Disc type & Association \\
-- & (deg) & (deg) & -- & -- \\
\hline
\endhead
\hline
\endfoot
\multicolumn{6}{l}{Spectral types are taken from: (1) \cite{Torres2006}, (2) \cite{Houk1975},(3) \cite{Houk1988},}\\
\multicolumn{6}{l}{(4) \cite{Gray2006}, (5) \cite{Harlan1974}, (6) \cite{Luhman2010}, (7) \cite{Mohanty2013}, (8) \cite{Mora2001},}\\
\multicolumn{6}{l}{(9) \cite{Manoj2006}, (10) \cite{Meeus2013}, (11) \cite{Houk1999}, (12) \cite{Vieira2003}, }\\
\multicolumn{6}{l}{(13) \cite{Zuckerman2004}, (14) \cite{Webb1999}, (15) \cite{Rydgren1980}, (16) \cite{Furlan2009}, }\\
\multicolumn{6}{l}{(17) \cite{Vacca2011}, (18) \cite{Luhman2007}, (19) \cite{Luhman2004}, (20) \cite{deLaReza1989}, }\\
\multicolumn{6}{l}{(21) \cite{Gregorio-Hetem1992}, (22) \cite{Houk1978}, (23) \cite{Levenhagen2006}, (24) \cite{Weinberger2004}, }\\
\multicolumn{6}{l}{(25) \cite{Houk1982}, (26) \cite{Spezzi2008}, (27) \cite{Fedele2013}, (28) \cite{Bouy2009}, }\\
\multicolumn{6}{l}{(29) \cite{Alcala2014}, (30) \cite{Lopez2011}, (31) \cite{Carpenter2006}, (32) \cite{Kohler2000}, }\\
\multicolumn{6}{l}{(33) \cite{Kraus2008}, (34) \cite{Dent2013}, (35) \cite{Hughes1994}, (36) \cite{Mortier2011},}\\
\multicolumn{6}{l}{(37) \cite{Prato2003}, (38) \cite{Preibisch1998}, (39) \cite{Preibisch2002}, (40) \cite{Wichmann1997},}\\
\multicolumn{6}{l}{(41) \cite{McClure2010}, (42) \cite{Bouvier1992}, (43) \cite{Ricci2010}, (44) \cite{Erickson2011},}\\
\multicolumn{6}{l}{(45) \cite{Martin1998}, (46) \cite{Nilsson2009}, (47) \cite{Hales2014}, (48) \cite{Zuckerman2004b}}\\
\endlastfoot
\object{HD~105} & 1.468958 & -41.753067 & G0$^{1}$ & DD & TucHor \\  
\object{HD~377} & 2.107291 & 6.616806 & G2$^{1}$ & DD & - \\  
\object{HD~3003} & 8.182958 & -63.0315 & A0$^{2}$  & DD & TucHor \\  
\object{HD~3670} & 9.736267 & -52.534284 & F5$^{2}$ & DD & - \\  
\object{HD~9672} & 23.657412 & -15.676359 & A1$^{3}$ & DD & Argus \\  
\object{tau~Ceti} & 26.017 & -15.937472 & G8.5$^{4}$ & DD & - \\  
\object{W3~IRS5} & 36.419125 & 62.097361 & - & 0 & - \\  
\object{HD~15115} & 36.567667 & 6.292556 & F4$^{5}$  & DD & - \\
\hline
  \end{longtable}
}

\onllongtab{2}{
\begin{longtable}{llllccc}
\caption{\label{YSO_fluxes} Fluxes at 63 $\rm \mu m$ for YSOs in the sample}\\
\hline\hline
Source name & RA & Dec & obs ID & $\rm F_ {[OI]}$ & $\rm F_{H {2}O}$ & $\rm F_{63\mu m}$ \\
-- & (deg) & (deg) & -- & ($\rm 10^{-18}W/m^{2}$) & ($\rm 10^{-18}W/m^{2}$) & (Jy) \\
\hline
\endfirsthead
\caption{continued.}\\
\hline\hline
Source name & RA & Dec & obs ID & $\rm F _{[OI]}$ & $\rm F_{H {2}O}$ & $\rm F_{63\mu m}$   \\
-- & (deg) & (deg) & -- & ($\rm 10^{ -18}W/m^{2}$) & ($\rm 10^{ -18}W/m^{2}$) & (Jy) \\
\hline
\endhead
\hline
\endfoot
\hline
\multicolumn{6}{l}{$\rm ^{*}$: observation was mis-pointed.}\\
\multicolumn{6}{l}{$\rm ^{**}$: source is located outside the central spaxel.}\\
\multicolumn{6}{l}{$\rm ^{***}$: the source shows absorption in the central spaxel.}
\endlastfoot
HD   105 & 1.468958 & -41.753 & 1342199239 & $\rm < 6$ & $\rm < 6$ & $\rm < 0.5$ \\ 
HD 377 & 2.107292 & 6.616806 & 1342212530 & $\rm < 7$ & $\rm < 7$ & $\rm < 0.5$ \\ 
HD   3003 & 8.182958 & -63.0315 & 1342199240 & $\rm < 7$ & $\rm < 7$ & $\rm < 1.0$ \\ 
HD  3670 & 9.736267 & -52.5342 & 1342269870 & $\rm < 7$ & $\rm < 7$ & $\rm < 0.5$ \\ 
HD   9672 & 23.65741 & 5.6763 & 1342188424 & $\rm < 11$ & $\rm < 11$ & $\rm 1.9 \pm 0.4$ \\ 
tau Ceti & 26.017 & 5.9374 & 1342189611 & $\rm < 8$ & $\rm < 8$ & $\rm < 0.624$ \\ 
W3 IRS5 & 36.41916 & 62.0975 & 1342229091 & $\rm 22026 \pm 520$ & $\rm < 632$ & $\rm 6958 \pm 20$ \\ 
W3 IRS5 & 36.41916 & 62.0975 & 1342229093 & $\rm 25543 \pm 3329$ & $\rm < 1000$ & $\rm 8461 \pm 61$ \\
\end{longtable}
}

\onllongtab{3}{
\begin{longtable}{llcccc}
\caption{\label{Tab:YSO_fluxes_ext} [OI] fluxes at 63 $\rm \mu m$ from central spaxel, central 3x3 spaxels and IFU integrated }\\
\hline\hline
Source name & obs ID & $\rm F _{[OI]}$ &   $\rm F _{[OI]3x3}$ &  $\rm F _{[OI]5x5}$ & Ext?$\rm ^{*}$ \\
-- & -- & ($\rm 10^{ -18}W/m^{2}$) & ($\rm 10^{ -18}W/m^{2}$) & ($\rm 10^{ -18}W/m^{2}$) & --  \\
\hline
\endfirsthead
\caption{continued.}\\
\hline\hline
Source name & obs ID & $\rm F _{[OI]}$ &   $\rm F _{[OI]3x3}$ &  $\rm F _{[OI]5x5}$ & Ext?$\rm ^{*}$ \\
-- & -- & ($\rm 10^{ -18}W/m^{2}$) & ($\rm 10^{ -18}W/m^{2}$) & ($\rm 10^{ -18}W/m^{2}$) & -- \\
\hline
\endhead
\hline 
\endfoot
W3IRS5 & 1342229093 & 25543.0$\rm \pm$3328.9 & 373064.6$\rm \pm$28956.2 & 1027951.6$\rm \pm$34137.0 & YYY \\
              & 1342229091 & 22026.4$\rm \pm$520.2 & 304283.8$\rm \pm$1250.4 & 728668.6$\rm \pm$1775.3 & YYY \\
SMM J032537+30451 & 1342263508 & 1668.3$\rm \pm$8.2 & 2260.4$\rm \pm$28.9 & 2368.6$\rm \pm$57.8 & YYN \\
LDN1448N & 1342263506 & 1609.5$\rm \pm$10.0 & 4001.2$\rm \pm$30.0 & 4927.1$\rm \pm$66.8 & YYY \\
L1448-C(S) & 1342263510 & 545.1$\rm \pm$12.6 & 1054.2$\rm \pm$31.3 & 1418.6$\rm \pm$95.2 & YYY \\
IRAS 03235+3004 & 1342264250 & 238.0$\rm \pm$11.0 & 361.0$\rm \pm$25.8 & 492.8$\rm \pm$72.9 & YYN \\
LDN 1455 & 1342204122 & 65.0$\rm \pm$14.0 & $\rm <$37.9 & $\rm <$102.2 & NNN \\
2MASSJ03283706+3113310 & 1342264248 & 187.8$\rm \pm$8.6 & 134.6$\rm \pm$17.8 & $\rm <$99.0 & NNN \\
\end{longtable}
\tablefoot{(*): the extended emission column shows the results for the 3x3 spaxels versus 1 spaxel/ 5x5 versus 1 spaxel and / 5x5 versus 3x3 spaxels. \\
}
}

\begin{table*}[!t]
\scriptsize
\caption{Fitted parameters for [OI] lines with multiple components}             
\label{Tab:multiG}      
\centering          
\begin{tabular}{lllllllllllll}     
\hline \hline 
Source& peak$\rm_{0}$ & $\rm \lambda_{0,0}$  & $\rm v_{0}$ & $\rm \sigma_{0}$ &   peak$\rm_{1}$ & $\rm \lambda_{0,1}$  & $\rm v_{1}$  & $\rm \sigma_{1}$ & peak$\rm_{2}$ & $\rm \lambda_{0,2}$  & $\rm v_{2}$  & $\rm \sigma_{2}$ \\
           & (Jy)                  & ($\rm \mu m$)   & (km/s)        &  ($\rm \mu m$)  & (Jy)                  & ($\rm \mu m$)    & (km/s)                  &  ($\rm \mu m$) & (Jy)                  & ($\rm \mu m$)        & (km/s)              &  ($\rm \mu m$)    \\
\hline
T Tau  & 15.52 & 63.159 & -123 & 0.0074 & 309.76 & 63.182 & -14 & 0.0075 & 209.03 & 63.19 & 24  & 0.0101 \\
FS Tau$\rm ^{*}$  & 16.15 & 63.183 & -10 & 0.007 & 7.78 & 63.189 & 19 & 0.0122 & 0.0 & 0.0 & 0.0 & 0.0 \\
DG Tau$\rm ^{*}$& 17.62 & 63.181 & -20 & 0.0072 & 14.3 & 63.182 & -13 & 0.0149 & 0.0 & 0.0 & 0.0 & 0.0 \\ 
HL Tau$\rm ^{*}$  & 25.11 & 63.197 & 55 & 0.0083 & 3.92 & 63.185 & 0.0 & 0.0279 & 0.0 & 0.0 & 0.0 & 0.0 \\ 
TMR 1 & 2.58 & 63.156 & -189 & 0.012 & 10.36 & 63.203 & 85  & 0.0185 & 0.0 & 0.0 & 0.0 & 0.0\\
UY Aur$\rm ^{*}$ & 12.77 & 63.183 & -10 & 0.0065 & 8.03 & 63.188 & 16 & 0.0133 & 0.0 & 0.0 & 0.0 & 0.0 \\
\hline                  
\end{tabular}
\tablefoot{(*): the velocity separation between the components is smaller than the spectral resolution ($\rm \sim 88~km/s$).}
\end{table*}

\acknowledgements
The authors would like to thank the anonymous referee for a very fruitful discussion that helped to improve the quality of the paper. P.R.M. acknowledges funding from the ESA Research Fellowship program. C.E. and B.M. are partly supported by Spanish Grant AYA 2014-55840-P.

\bibliographystyle{aa} 
\bibliography{biblio.bib}

\appendix

\section{Line emission maps}

\begin{figure*}[!t]
\centering
\includegraphics[]{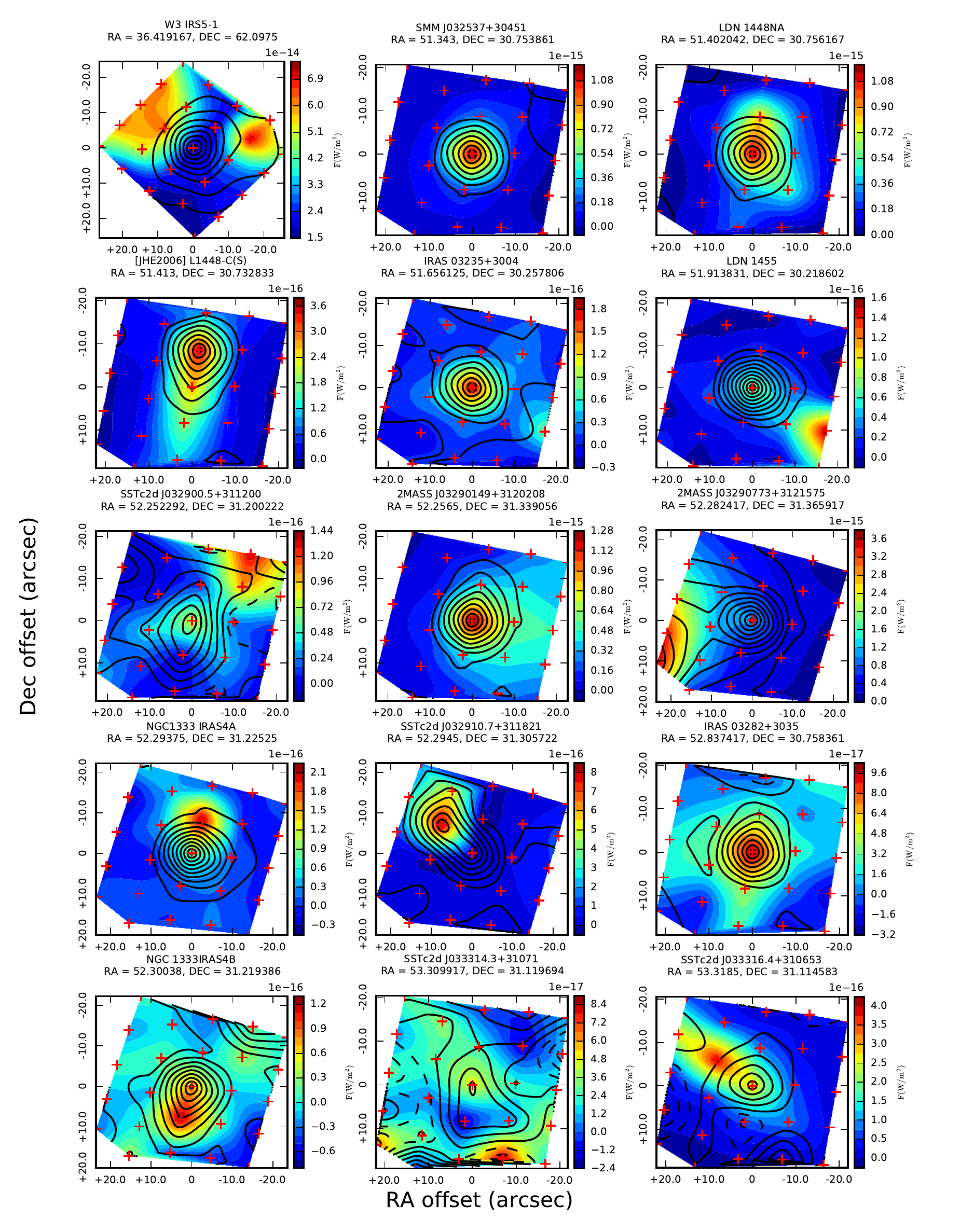}
\caption{[OI]  line emission at 63 $\rm \mu m$(coloured contours) and 63 $\rm \mu m$ continuum contours (solid black lines) for sources identified as extended by any of the three tests used. The positions of spaxels are marked with red plus signs. }
   \label{Fig:lineMaps}
\end{figure*}

\addtocounter{figure}{-1}
\begin{figure*}[!t]
\centering
\includegraphics[]{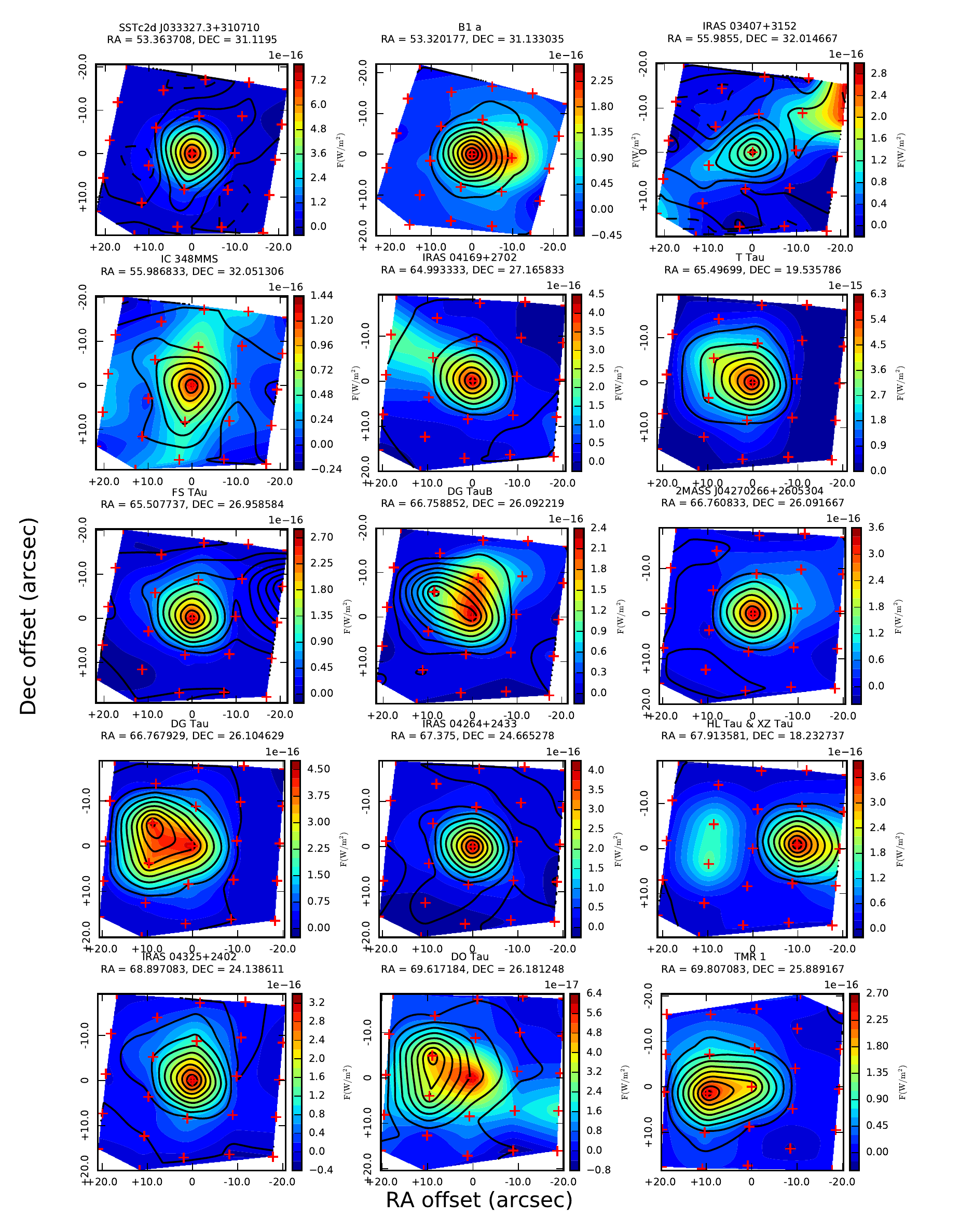}
\caption{continued.}
   \label{Fig:lineMaps1}
\end{figure*}

\addtocounter{figure}{-1}
\begin{figure*}[!t]
\centering
\includegraphics[]{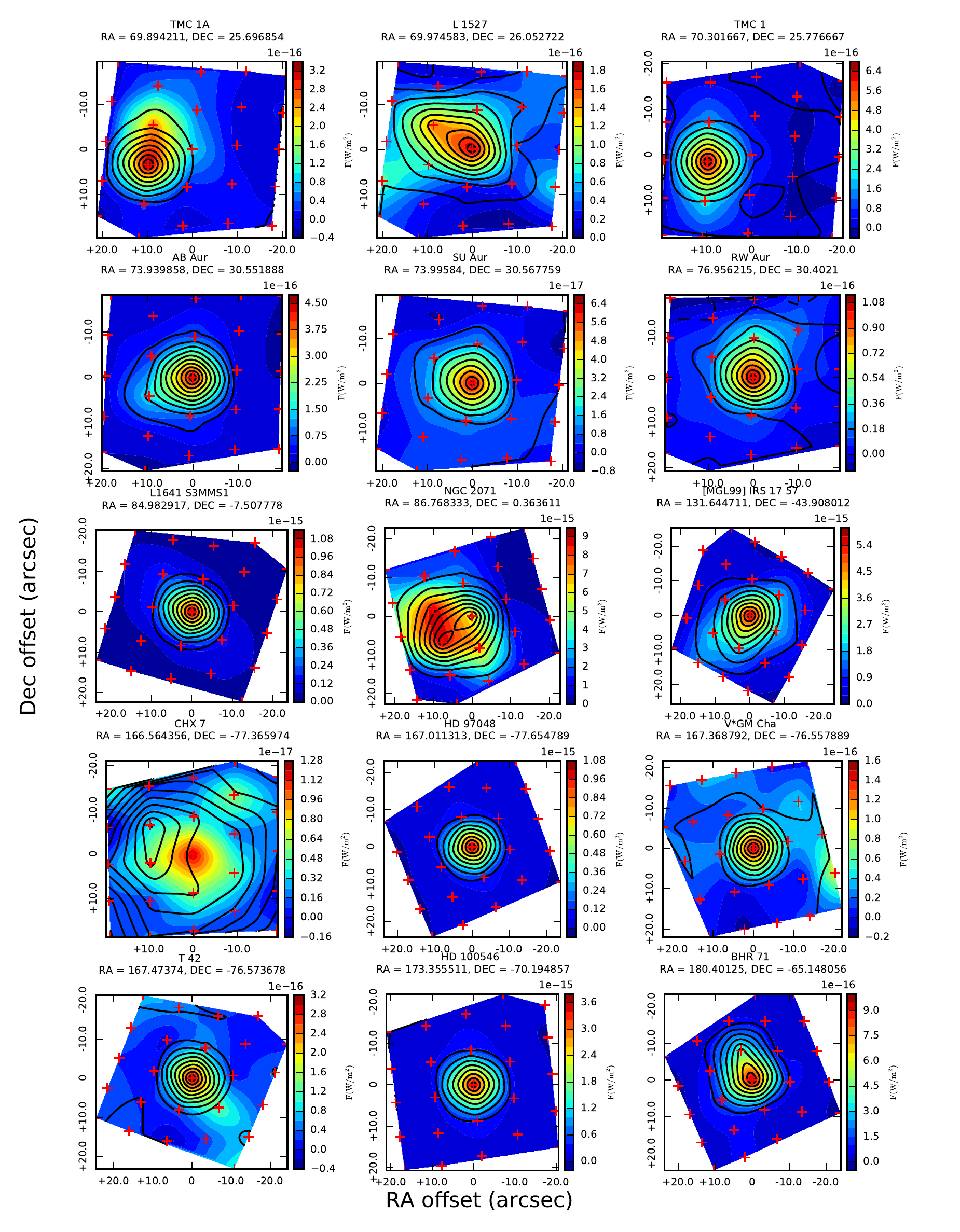}
\caption{continued.}
   \label{Fig:lineMaps2}
\end{figure*}

\addtocounter{figure}{-1}
\begin{figure*}[!t]
\centering
\includegraphics[]{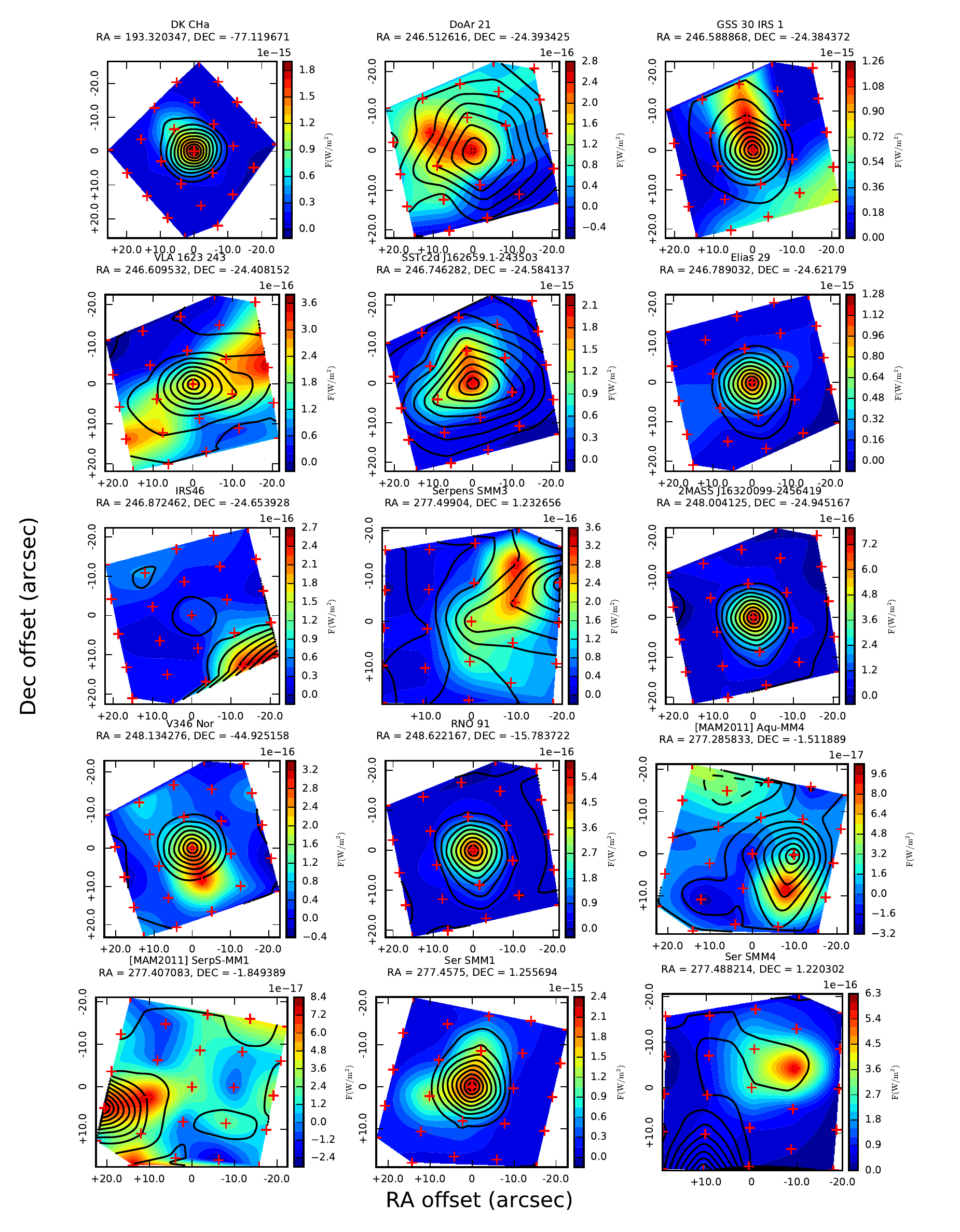}
\caption{continued.}
   \label{Fig:lineMaps3}
\end{figure*}

\addtocounter{figure}{-1}
\begin{figure*}[!t]
\centering
\includegraphics[]{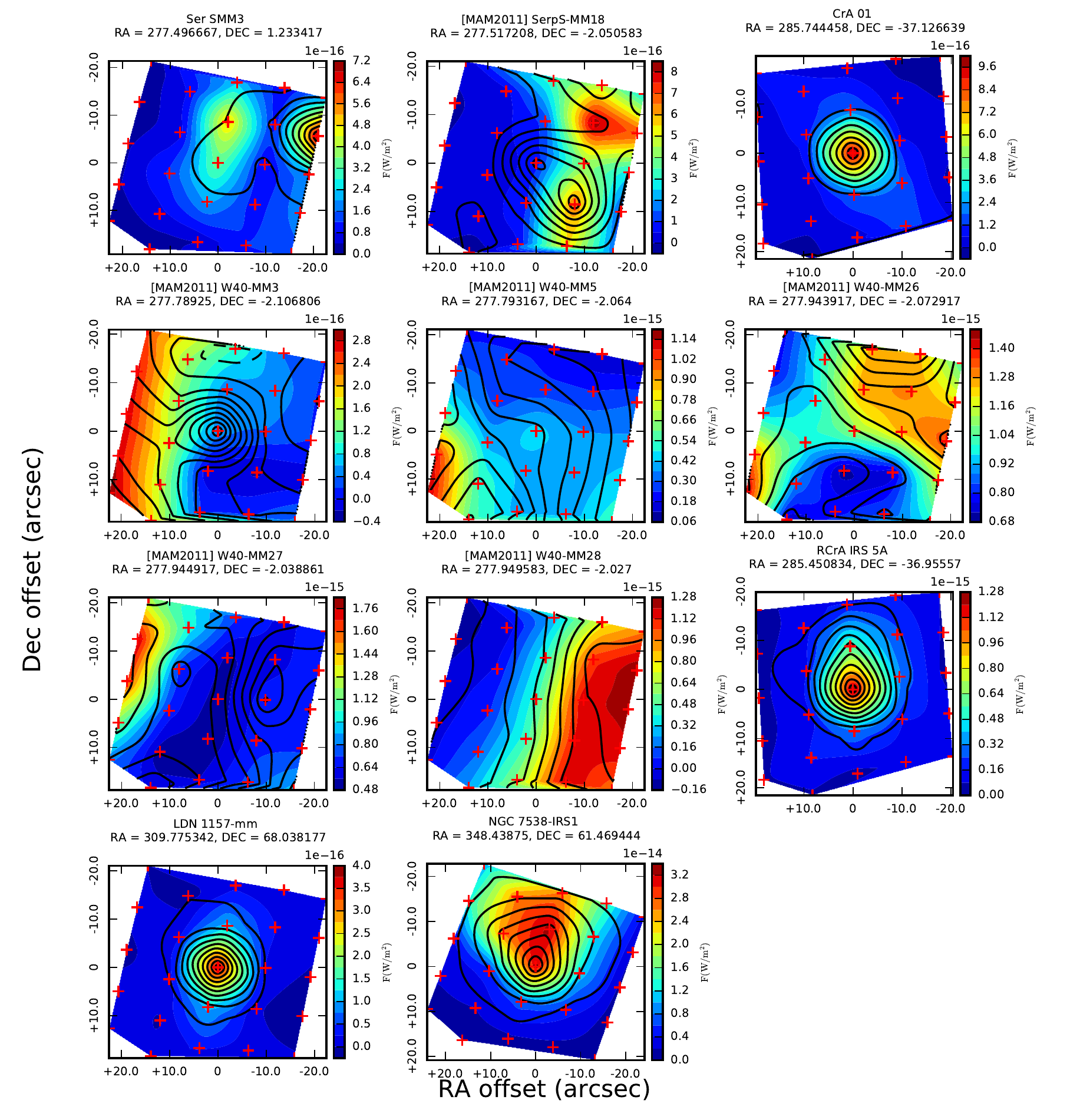}
\caption{continued.}
   \label{Fig:lineMaps4}
\end{figure*}

\section{Residual line emission maps}

\begin{figure*}[!t]
\centering
\includegraphics[]{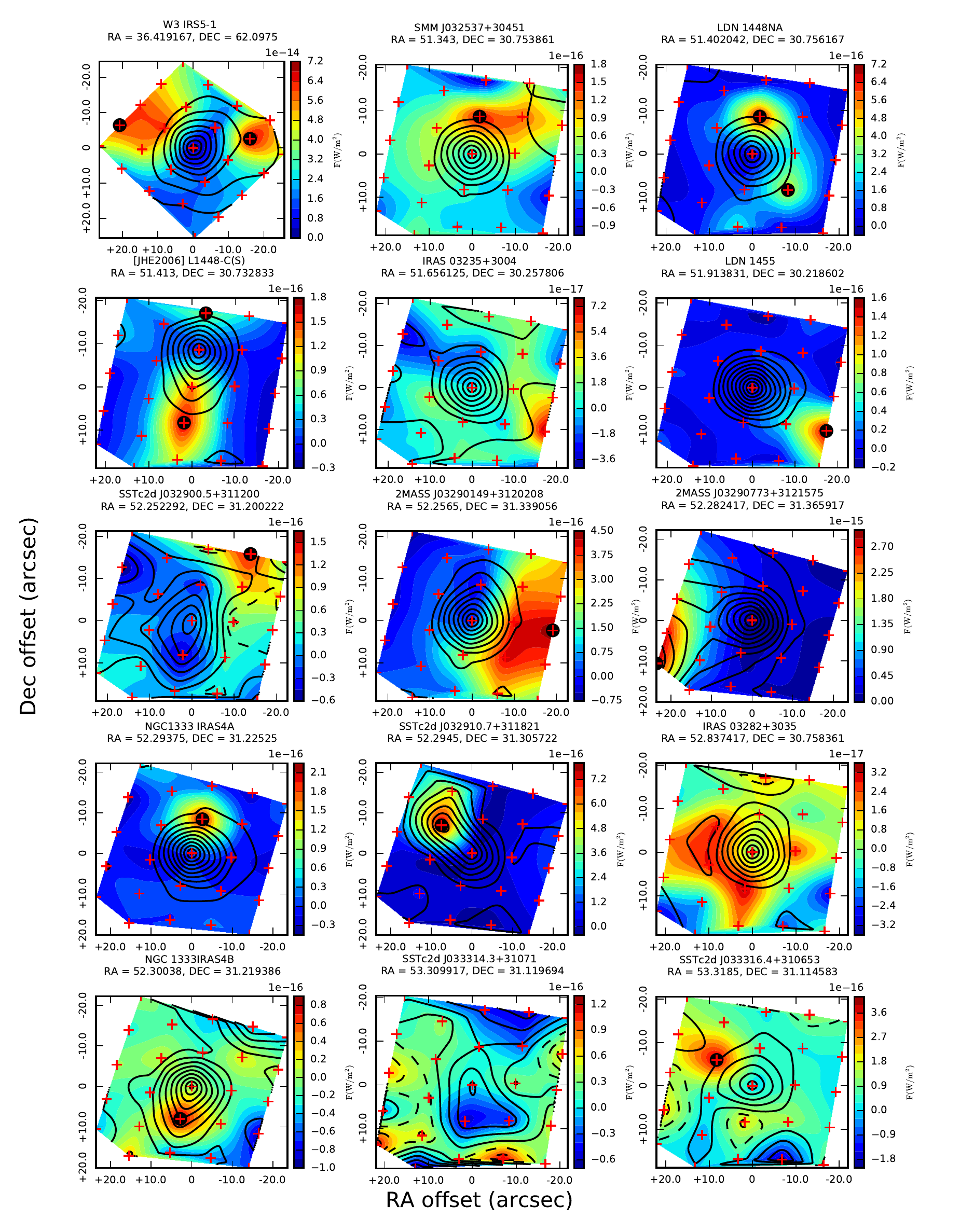}
\caption{[OI] residual line emission at 63 $\rm \mu m$(coloured contours) and 63 $\rm \mu m$ continuum contours (solid black lines) for sources identified as extended by any of the three tests used. The positions of spaxels are marked with red plus signs. Spaxel with 5$\sigma$ residual detections are surrounded by black dots.}
   \label{Fig:ResidualMaps}
\end{figure*}

\addtocounter{figure}{-1}
\begin{figure*}[!t]
\centering
\includegraphics[]{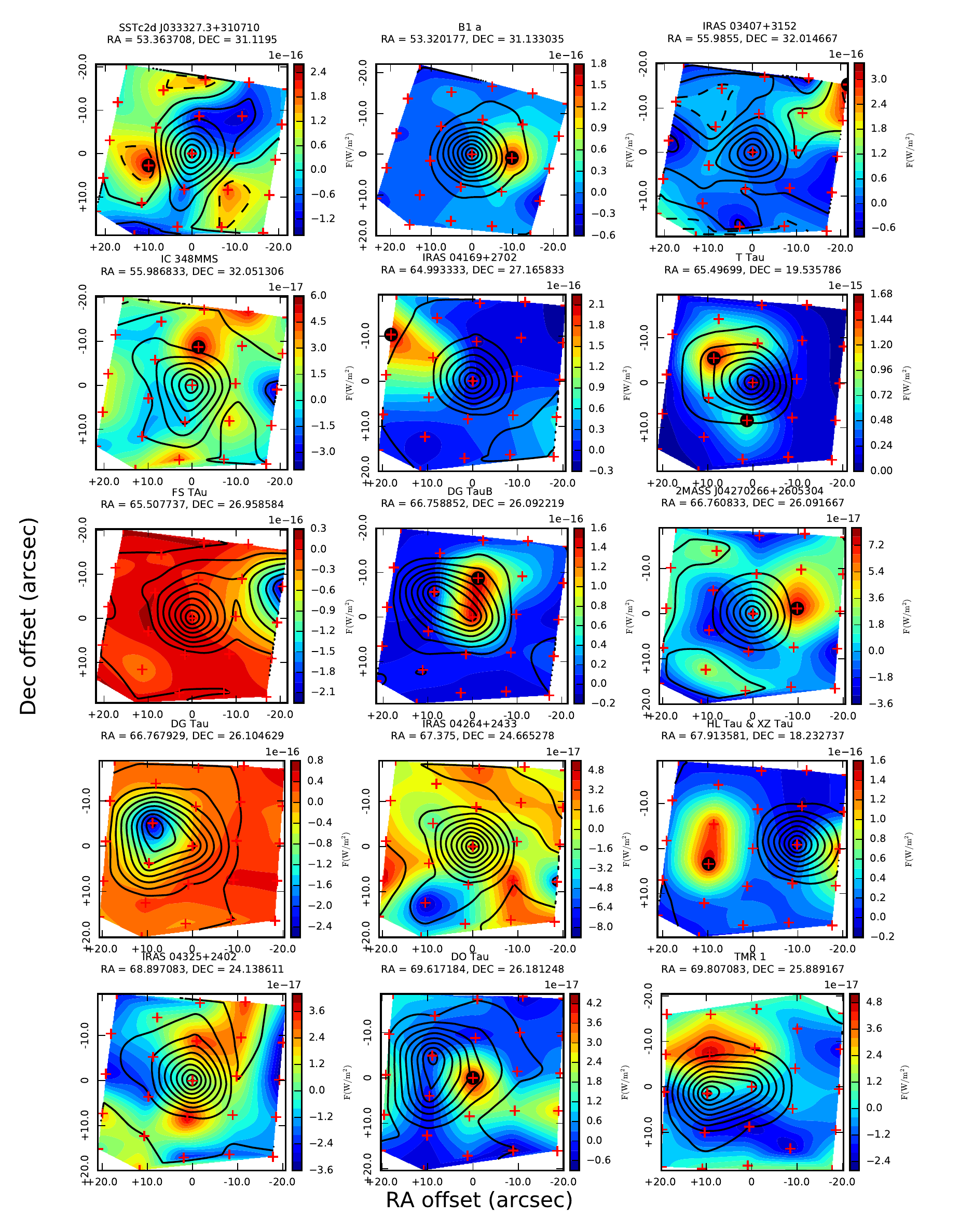}
\caption{continued.}
   \label{Fig:ResidualMaps1}
\end{figure*}

\addtocounter{figure}{-1}
\begin{figure*}[!t]
\centering
\includegraphics[]{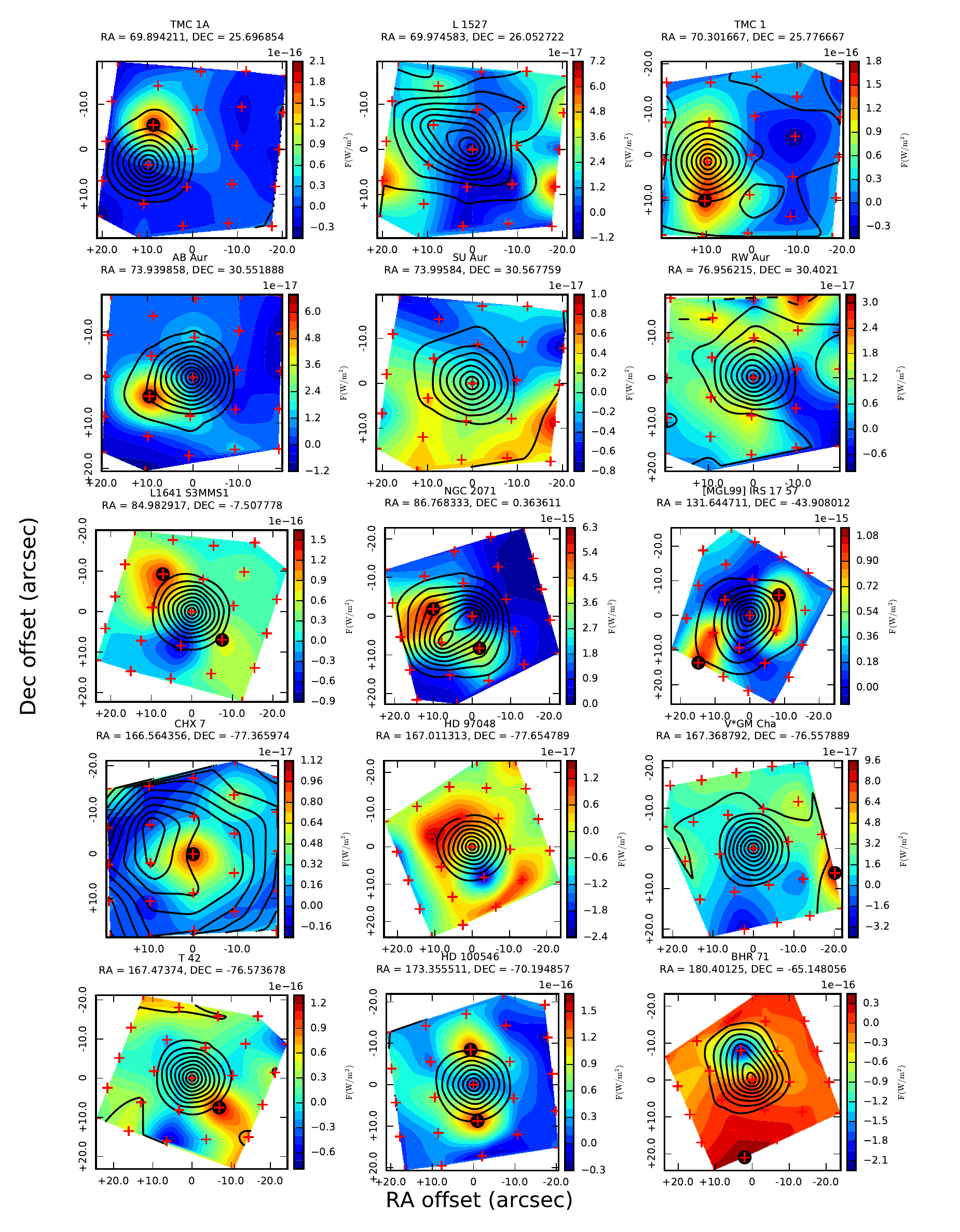}
\caption{continued.}
   \label{Fig:ResidualMaps2}
\end{figure*}

\addtocounter{figure}{-1}
\begin{figure*}[!t]
\centering
\includegraphics[]{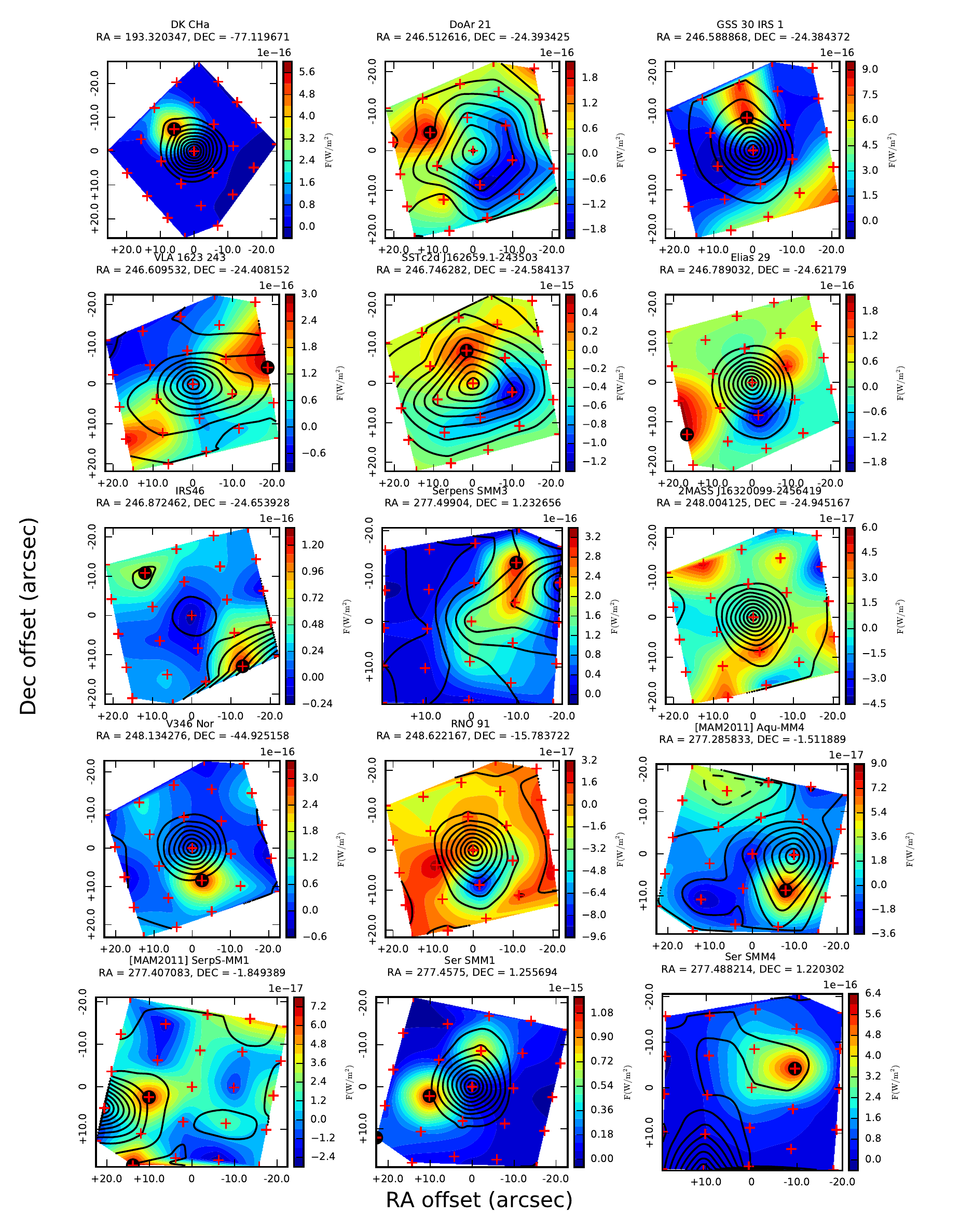}
\caption{continued.}
   \label{Fig:ResidualMaps3}
\end{figure*}

\addtocounter{figure}{-1}
\begin{figure*}[!t]
\centering
\includegraphics[]{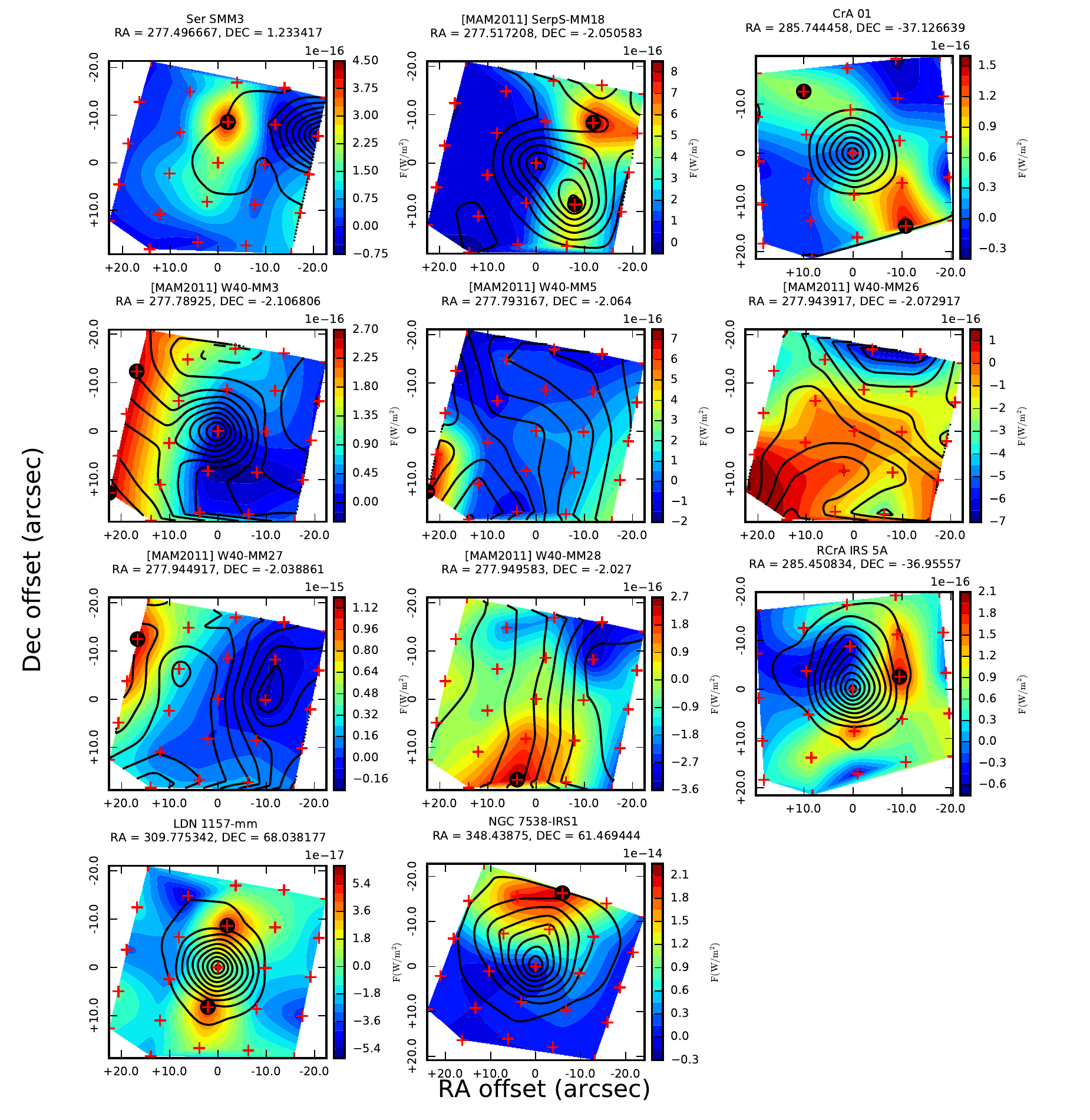}
\caption{continued.}   
\label{Fig:ResidualMaps4}
\end{figure*}

\end{document}